\documentclass[12pt,a4paper]{article}

\usepackage{amsthm,amsmath,amsfonts,amssymb}
\usepackage{times,latexsym,dsfont}
\usepackage{eurosym} %%For Euro symbol.

\usepackage{subfig} %% %For multiple figures in one panel
\usepackage{graphicx} %%
\usepackage[countmax]{subfloat} %% %For multiple figures in one panel

\usepackage{nohyperref}
\usepackage{xcolor}
\hypersetup{
    colorlinks,
    linkcolor={red!50!black},
    citecolor={blue!50!black},
    urlcolor={blue!80!black}
}

\usepackage{lscape} %%
\usepackage{listings} %%
\usepackage{setspace} %%
\usepackage{url}      %%
\usepackage{makeidx}  %% %Verzeichnisse vorsortieren
\usepackage{rotating,here} %%

\usepackage{lineno} %%Zeilennummern

\usepackage{tabularx} %% %For multiple panels in tables
\usepackage{booktabs} %%
\newcolumntype{Y}{>{\raggedleft\arraybackslash}X} %%

\usepackage{color}
\usepackage{psfrag}
\usepackage{lmodern}
\usepackage[USenglish]{babel}
\usepackage[T1]{fontenc}
\usepackage[latin1]{inputenc}
\theoremstyle{plain}

\theoremstyle{definition}

\usepackage{tikz,color}
\usepackage{pgflibraryshapes}
\usetikzlibrary{trees,arrows, decorations.markings, positioning}
\definecolor{grau}{rgb}{0.9,0.9,0.9}
\tikzstyle{vecArrow} = [line width=2mm, draw=black, -triangle 60,
                        postaction = {draw, line width=5mm, shorten >=6mm, -},
                        preaction = {decorate}]
\tikzstyle{innerWhite} = [line width=1.9mm, draw=grau, -triangle 60,
                        postaction = {draw, line width=4.5mm, shorten >=5.5mm, -}]

\begin{document}

\renewcommand{\thefootnote}{\arabic{footnote}}

\begin{center}
\large{\bf Short-Term Wind Speed Forecasting in Germany\\}
\vspace{10mm} \normalsize \today
\small

Daniel Ambach$^1$

\end{center}

\thispagestyle{empty} \small

\normalsize

\begin{abstract}
\noindent The importance of renewable power production is a set goal in terms of the energy turnaround. Developing short-term wind speed forecasting improvements might increase the profitability of wind power. This article compares two novel approaches to model and predict wind speed. Both approaches incorporate periodic interactions, whereas the first model uses Fourier series to model the periodicity. The second model takes $l_{2,p}$ generalised trigonometric functions into consideration. The aforementioned Fourier series are special types of the p-generalised trigonometrical function and therefore model 1 is nested in model 2. The two models use an ARFIMA-APARCH process to cover the autocorrelation and the heteroscedasticity. A data set which consist of 10 minute data collected at four stations at the German-Polish border from August 2007 to December 2012 is analysed. The most important finding is an enhancement of the forecasting accuracy up to three hours that is directly related to our new short-term forecasting model.
\end{abstract}

\begin{small}{\bf Keywords:} Time series regression; generalised trigonometric functions; ARFIMA-APARCH process; high-frequency data; wind speed forecasts.
\end{small}

\bigskip

\begin{bfseries}
Addresses:
\end{bfseries}

\begin{small}
$^{1}$ \textbf{Corresponding Author:} Daniel Ambach, European University Via\-dri\-na, Chair of Quantitative Methods and Statistics,
Post Box 1786, 15207 Frankfurt (Oder), Germany, Tel. +49 (0)335 5534 2983, Fax +49 (0)335 5534 2233, ambach@europa-uni.de.\vspace{0.1cm}
\end{small}
\pagestyle{plain}
\setcounter{page}{1}
\onehalfspacing

%Introduction

\section{Introduction}

Fossil fuels like oil, gas and coal are rare resources and it is well known that they are finite. The catastrophe in Fukushima (Japan) put a negative light onto nuclear energy and forced people to reconsider nuclear power plants. One of the most emergent renewable energy sources is the wind power. \cite{WWEA2014} report that the worldwide wind capacity grew by $5.5\%$ from the end of 2013 to June 2014. Moreover, they state that this capacity generates approximately $4\%$ of the worldwide electricity supply. The electricity sector of the European Union (EU) tends towards wind energy, as \cite{europeanwind} remark. Besides, they observe that within the last 15 years, $28\%$ of the newly installed power production capacity stem from wind power. The leading wind power country in the European Union is Germany, which has an installed capacity of $36.488$ MW by June 2014 \cite{WWEA2014}. \cite{europeanwind} find that fifteen EU countries have an installed capacity of wind power of each $>1.000$ MW. Poland and Romania are more recent EU members which also exceed the threshold of $1.000$ MW \cite{europeanwind}. The installed capacity in Poland amounts to $3.727$ MW \cite{WWEA2014}. The geographical region at the German-Polish border is very much alike, but the installed capacity varies significantly, as \cite{WWEA2014} report.

The profit of wind power strongly depends on several properties, as \cite{burton2001wind} point out, but the focus of this article is the wind speed prediction. The wind speed is an essential part of the wind power. Moreover, accurate wind forecasts can be used to optimise the overall electricity supply, particularly the feed-in of wind power. Developing new wind forecasting models depends on the previously specified purpose and forecasting horizon. \cite{soman2010review}, \cite{zhu2012short} as well as \cite{jung2014current} provide a broad review of differing wind speed and wind power prediction models. As \cite{jung2014current} point out, conventional time series models are easy to formulate and often used for short-term predictions. The wind speed possesses different properties like periodicity, persistence, heteroscedasticity as well as further characteristics. \cite{Haslett1989} show a wind speed forecasting approach which contains an autoregressive fractionally integrated moving average (ARFIMA) process. It is one of the first models that comprise a long memory process. Since their contribution, the ARFIMA model has become very common to model the wind. Hence, \cite{Taylor2009} examine extensions of the approach. They compare an ARFIMA model with generalised autoregressive conditional heteroscedasticity (GARCH) and periodic regressors to calibrated and smoothed ensemble-based wind forecasts for hourly data. The periodic behaviour of wind speed depends on the temporal resolution of the underlying data set. Frequently, such periodicities are modelled by Fourier series \cite{Taylor2009,benth2010analysis}. Altogether, miscellaneous articles report that wind speed is autocorrelated and shows periodic behaviour. The wind speed characteristics are commonly combined in several ways. \cite{jiang2013very} implement a model which is based on Bayesian theory and structural breaks to implicate these properties. \cite{Ambach2015} cover a comparison of different long range dependence models with time varying regressors and different estimation methods. A further characteristic is given by the heteroscedasticity of wind. \cite{benth2010analysis} use a combination of ARMA models with periodic regressors. Furthermore, they introduce Fourier series for the variance part to model the heteroscedasticity. \cite{Ambach2015} find that their data set contains asymmetric conditional heteroscedasticity. Certainly, this is only a brief overview of prevalent wind characteristics. \cite{Hering2010} incorporate spatial information within their forecasting models to extend the accuracy. They perform a comparison of a regime switching model and a bivariate skew-t model. \cite{zhu2014space} develop a  generalisation of the regime switching approach which is introduced by \cite{gneiting2006calibrated}. The generalised regime switching space-time model includes several wind characteristics. Our model covers periodicity, persistence and heteroscedasticity and thus includes information to provide suitable short-term forecasts.

As \cite{zhu2012short} point out, those forecasting horizons are beneficial for dispatch planning and for economic load decisions. This article contains two wind speed prediction approaches. Both models use an autoregressive fractionally integrated moving average (ARFIMA) process with periodic regressors and conditional heteroscedasticity among the variance. Particularly, the two approaches incorporate an asymmetric power generalised autoregressive conditional heteroscedasticity (APARCH) process within the conditional variance \cite{Ding1993}. Moreover, both models take annual and diurnal interactions into account. The difference of the two approaches is linked to the periodic mean part in addition to the autocorrelation. The first model uses the classical sine and cosine Fourier series. The second model contains $l_{2,p}$ generalised trigonometric functions \cite[see][]{richter2007generalized} to model seasonal changes. The analysed data set consists of 10 minute wind speed observations. This data is used to fit the model and perform predictions up to three hours at four different stations at the German-Polish border. The forecasting results are evaluated by several accuracy measures. Furthermore, a loss function as introduced by \cite{Hering2010} is taken into consideration. The importance of this loss function is underlined by \cite{zhu2012short}.

In Section 2, the general model for our wind speed is introduced. Following the theoretical content, we discuss the estimation of the model. Section 3 depicts several methods to evaluate the forecasting accuracy of the selected models. Finally, Section 4 draws a conclusion of the results discussed in this paper.

%%%%%%%%%%%%%%%%%%%%%%%%%%%%%%%%%%%%%%%%%%%%%% SECTION2 %%%%%%%%%%%%%%%%%%%%%%%%%%%%%%%%%%%%%%%%%%%%%%%%%%%%%%%%%%%%%%%

\section{Description of two wind speed models and their in-sample properties}

The contemplated wind speed data set is recorded at four stations in the state of Brandenburg. We consider high-frequency data (10 minute) which is provided by the ``Deutscher Wetterdienst'' (DWD). Brandenburg is Germany's fifth biggest territorial area. The stations are situated near the Polish border in Manschnow ($52^\circ 33'N$ $14^\circ 32'E $), Lindenberg ($52^\circ 13'N$ $14^\circ 07'E $), Angerm\"unde ($53^\circ 02'N$ $14^\circ 00'E $) and Gr\"unow ($53^\circ 19'N$ $13^\circ 56'E $). The wind speed data consists of approximately $280,000$ observations within a time frame from August 2007 to December 2012, with $0.1\%$ missing values. The gaps of missing data are not longer than a few hours in all cases. We take several interpolations into consideration, but they provide nearly identical results. Therefore, we decide to use linear interpolation, as it is the simplest and the least computationally intensive procedure. The in-sample model ca\-li\-bra\-tion starts in August 1, 2007 and reaches to December 31, 2011.

\begin{figure}
\begin{center}
\includegraphics[width=1\textwidth]{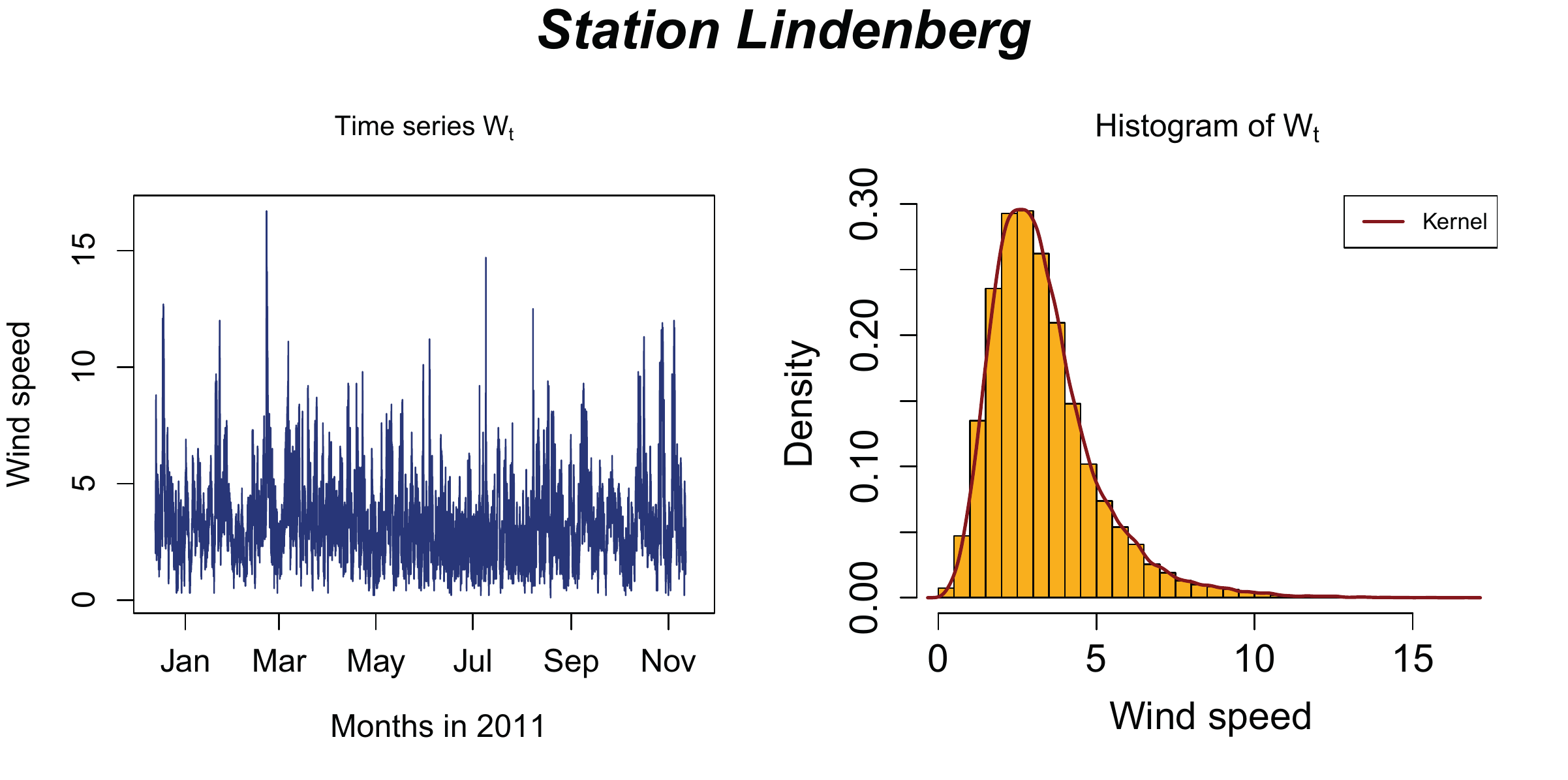}
\caption{Wind speed sample of Lindenberg reported in 2011 and corresponding histogram.}
\label{fig:ts}
\end{center}
\end{figure}

\begin{figure}
\begin{center}
\includegraphics[width=.6\textwidth]{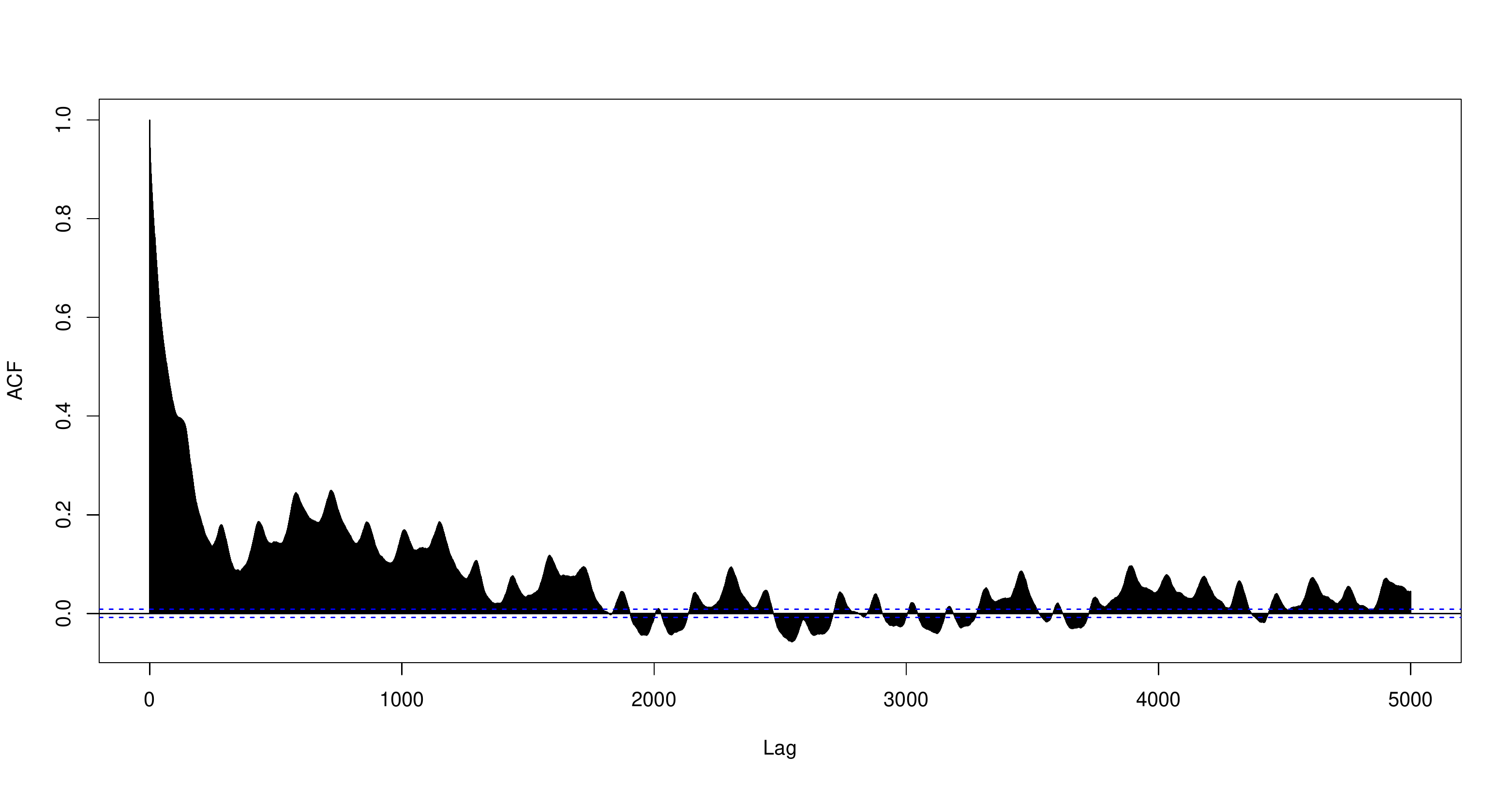}
\caption{Wind speed autocorrelation function plot of Lindenberg.}
\label{fig:acf}
\end{center}
\end{figure}

The considered high-frequency wind data exhibit several of the aforementioned wind speed characteristics. Figure \ref{fig:ts} shows a sample of the wind speed observations and the corresponding histogram of station Lindenberg. We observe that the wind is rapidly changing and involves a high correlation structure, as shown in Figure \ref{fig:acf}. This figure shows the autocorellation function up to lag $5,000$ which corresponds to more than one month. The other measurement stations provide a similar characteristics.

\begin{center}
  [Figure \ref{fig:ts} and \ref{fig:acf} about here.]
\end{center}

\begin{figure}
\begin{center}
\includegraphics[width=1\textwidth]{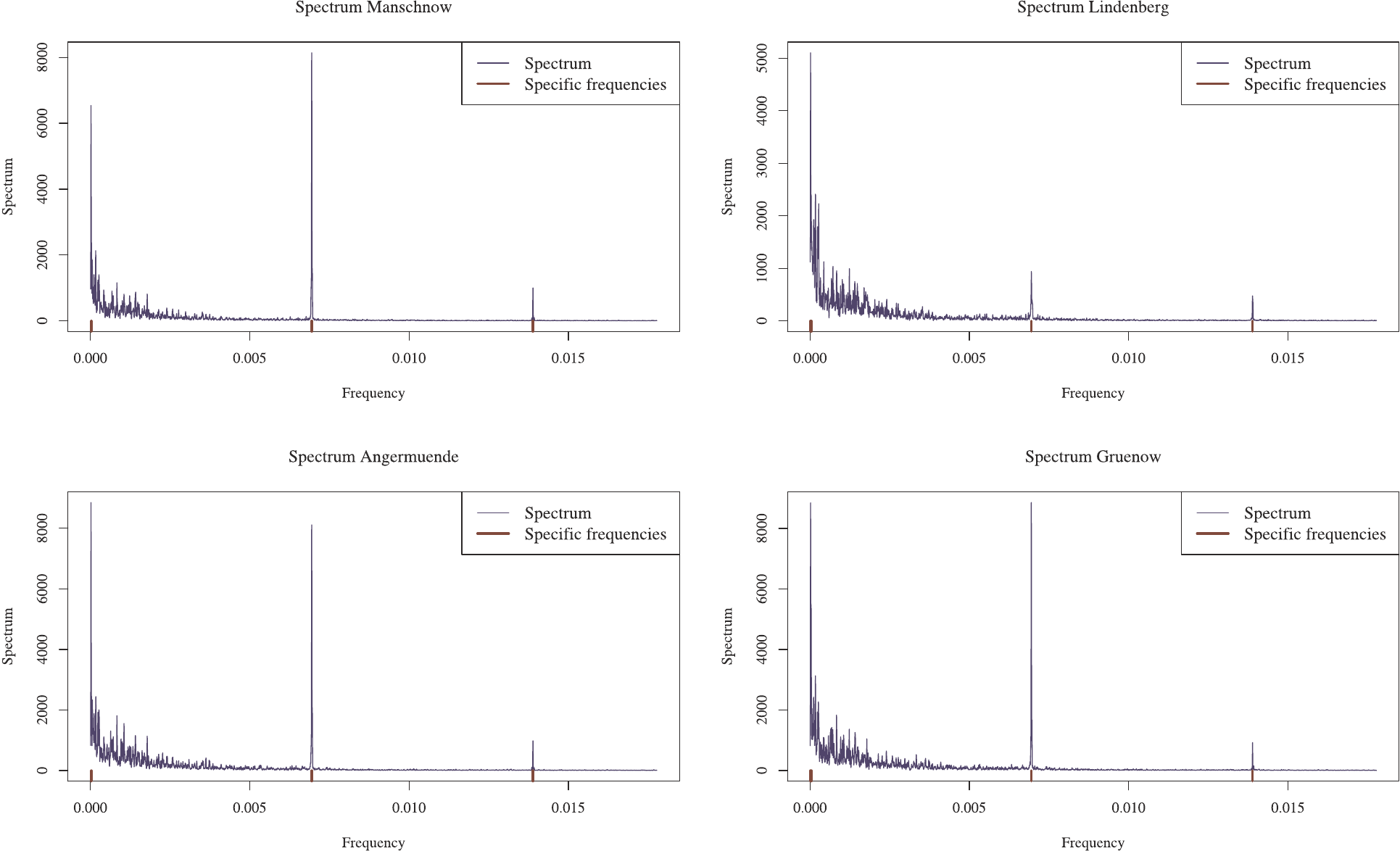}
\caption{Estimation of the spectral density at each single station.}
\label{fig:spectrum}
\end{center}
\end{figure}

\noindent The periodic characteristic of wind depends on the frequency of the observed data. In particular, we analyse the spectral density to find all necessary periods. The basic tool for these inferences is the periodogram, which is supposed to be the sample analogue of the spectrum. A detailed description of this is given by \cite{Brockwell2009}. Figure \ref{fig:spectrum} shows the smoothed periodogram by using the Fast Fourier Transformation (FFT). The peaks included in this picture clearly depict a periodic behaviour for different frequencies $\omega_N$. In general, we observe $s_1, \ldots , s_N \in \mathds{R}$ periodicities with $s_1  > \ldots > s_N$ and $s_N = 1/\omega_N$. Our choice of the peak depends on the period $s \in \{s_1, \ldots, s_N\}$ on the one hand and on the total value of the estimated spectrum on the other hand. Near the point of origin we detect periods of one year and more. Bearing in mind that we analyse a data set of about five and a half years, we neglect peaks with periodic cycles of more than one year. Altogether, we consider at least four periodicities. The selected peaks correspond to diurnal $s_1 =1/.000019= 52560$ and annual $s_2 =1/.00694 = 144$ periodicities. In addition, we consider two sub-sets of the previous determined frequencies, i.e. half-day  ($s_1/2 = 1/.0138$) and half-year ($s_2/2 = 1/.000038$) periodicity. Such frequencies reflect the day-night cycle and the summer and winter term.

\begin{center}
  [Figure \ref{fig:spectrum} about here.]
\end{center}

\noindent Thus, we observe persistence, periodicity and heteroscedasticity within our wind speed data similar to the approaches by \cite{Taylor2009}, \cite{benth2010analysis} and \cite{Haslett1989}. Apparently, these models use different data sets. For example \cite{Ewing2006} implement an ARMA-GARCH in mean model to their 15 minute wind data set. \cite{Taylor2009} use hourly and \cite{benth2010analysis} use three hour wind data. We apply the aforementioned approaches to our data, but they do not provide an overall satisfying result. For instance, \cite{benth2010analysis} take periodic regressors within the variance part into consideration, whereas our variance data shows an asymmetric conditional heteroscedasticity. Such a result is related to the underlying data set and therefore, we have to develop another approach. We provide two novel wind speed models which use a new method to capture the periodic behaviour of wind.

\subsection{Model description}

Let $\{W_{t}\}$ be a high-frequency process of wind speed, measured in $m/s$. The univariate time series of wind speed measurements is given by $W_{144(r-1)+o} = W_{t}$, where $r \in  \{1,...,N\}$ is the day, $o \in \{1,...,144\}$ is the 10 minute data point within a day. We are able to generalise the model \cite[see][]{Brockwell2009}

\begin{equation}
\label{general}
    W_{t} = Intercept_{t} + Trend_{t} + Seasonal_{t} + \epsilon_{t} \vspace*{0.3cm},
\end{equation}

\noindent where $\{\epsilon_{t}\}$ is the residual process with zero mean and time dependent variance $\{\sigma_t\}$. The residual process incorporates autoregressive coefficients and conditional heteroscedasticity within the variance. The general model (\ref{general}) is the most simplified expression of the wind speed time series. Hereafter, we explain the seasonal regressors in more detail.

\begin{eqnarray}
\label{equation:trend}
  Trend_{t}  +  Intercept_{t}     &=&  \vartheta_{trend} \cdot t + \vartheta_{11}, \\
\label{equation:intandseason}
  Seasonal_{t}    &=&  \sum^{k_1}_{i_1=1} \sum^{k_2}_{i_2=1} I_{i_1 i_2} \vartheta_{i_1 i_2} f_{i_1 ,i_2}^{s_1}(t) f_{i_1,i_2 }^{s_2}(t) , \vspace*{0.4cm}
\end{eqnarray}

\noindent where $i_1 = \{1,...,k_1\}$ and $i_2 = \{1,...,k_2\}$ provide up to $k_1$ and $k_2$ time dependent periodic functions.  $I_{i_1 \, i_2} \in \{ 0, 1 \}$ represents an element of the indicator matrix $\boldsymbol{I}$ with $k_1 \times k_2$ dimension and $\vartheta_{trend}$ is the trend coefficient which is constant over time. Moreover, the function set $f_{i}^{s} \in \{f_{i_1 ,i_2}^{s_1},f_{i_1,i_2 }^{s_2}\}$, where $i \in \{i_1,i_2\}$, provides periodic external variables given by

\begin{equation}
\label{equation:funct}
\ f_{i_1 i_2}^{s_1}(t) = \begin{cases} 1 &, \,  i_1 = 1 \\[.2cm]
 				\cos_{p}\left(\frac{\pi i_1 t }{s_1} \right) &, \,  i_1=\text{even}  \\[.2cm]
 				\sin_{p}\left(\frac{\pi (i_1-1) t }{s_1}\right) &, \,  i_1=\text{odd, } \backslash \{1\} \end{cases} \,
\end{equation}
and
\begin{equation}	
\label{equation:funct1}		
 \ f_{i_1 i_2}^{s_2}(t) = \begin{cases} 1 &, \,  i_2 = 1 \\[.2cm]
 				\cos_{p}\left(\frac{\pi i_2 t }{s_2} \right) &, \, i_2=\text{even}    \\[.2cm]
 				\sin_{p}\left(\frac{\pi (i_2-1) t }{s_2}\right) &, \, i_2=\text{odd, } \backslash \{1\} ,  \end{cases}  \, 				 \vspace*{0.3cm}
\end{equation}

\noindent where $\cos_{p}$ and $\sin_{p}$ are p-generalised periodic functions or Fourier, series if $p=2$. Commonly, we observe the classical way to model the periodicity by Fourier series, like \cite{Hering2010} and \cite{Taylor2009} show. Therefore, our first model uses the classical sine and cosine decomposition ($p=2$) as well. Our second model uses the p-generalised sine and cosine functions of \cite{richter2007generalized} ($p>0$) to obtain a more flexible periodic structure.  \cite{richter2011ellipses} mention that recently, the circle number $\pi$ is generalised for $l_{2,p}$ circles. Furthermore, \cite{richter2007generalized} use the p-generalised radius coordinates to define p-generalised trigonometrical functions for $p>0$. The p-generalised trigonometrical functions are given by \cite{richter2007generalized}

\begin{equation}
 \sin_{p}(\varphi) =	 \frac{\sin \varphi}{\left(\vert \sin \varphi \vert^{p}  + \vert \cos \varphi \vert^{p}\right)^{1/p}}
\end{equation}
and
\begin{equation}
 \cos_{p}(\varphi)= \frac{\cos \varphi}{\left(\vert \sin \varphi \vert^{p}  + \vert \cos \varphi \vert^{p}\right)^{1/p}} \, , \vspace*{0.3cm}
\end{equation}

\noindent where $\varphi \in \mathbb{R}$. Moreover, we have to describe the residual process $\epsilon_{t}$ which follows an ARFIMA-APARCH and is given by

\begin{eqnarray}
\label{equation:arfimaaparch}
\begin{array}{rcl}
\epsilon_{t} &\equiv & \phi (B) (1- B)^{d}  X_{t}  =  \theta (B) Z_{t} ,  \\[.2cm]
 & Z_{t}=& \sigma_{t} \eta_{t}  \qquad \text{where} \quad   \eta_{t} \overset{iid}{\sim} F_{\nu , \mu, \sigma, \xi }  \\[.2cm]
       &\sigma_{t}^\delta =&    \alpha_{0} + \sum\limits_{l=1}^{Q} \alpha_l (| Z_{t-l}| - \gamma_l Z_{t-l})^{\delta}  + \sum\limits_{m=1}^{P} \beta_m  \sigma_{t-m}^{\delta} , \vspace*{0.3cm}
 \end{array}
\end{eqnarray}

\noindent where $d \in (-0.5,0.5)$ is the differencing parameter, $B$ is the backward shift operator with $B^u X_{t} =  X_{t-u}$. The polynomials $\phi$ and $\theta$ are $\phi (B) = 1- \phi_1 B - \ldots - \phi_j B^j $ and $\theta (B) = 1+ \theta_1 B + \ldots + \theta_q B^q$. $\delta$ is the power of the APARCH process, $\gamma$ is the asymmetry parameter and $\{\eta_{t}\}$ are the residuals. \cite{Ding1993} point out that instead of the conditional variance, the conditional standard deviation with power $\delta$ follows an APARCH. Wind speed data is highly volatile and different wind regimes may lead to the asymmetric conditional heteroscedasticity. Thus, if the wind speed is relatively low, we observe a different conditional variance compared to high speed wind regimes. The residuals $\{\eta_{t}\}$ are described by a skew Student's t-distribution $F_{\nu, \mu, \sigma, \xi}$. One definition of this distribution is given by \cite{Fernandez1998}. They combine two halves of a symmetric t-distribution which are differently scaled. The skew-t distribution is given by

\begin{equation}
\label{equation:skewed-t}
              f_x(x) = \frac{2\xi}{(\xi^2+1)} \frac{\Gamma\big(\frac{\nu+1}{2}\big)}{\Gamma \big(\frac{\nu}{2}\big) \sqrt{\pi} \sqrt{\nu} \sigma} \bigg[ 1+ \frac{\big( \frac{x-\mu}{\sigma}\big)^2}{\nu} \bigg( \frac{1}{\xi^2} I (x \geq \mu) + \xi^2 I (x < \mu) \bigg)\bigg]^{-\frac{\nu+1}{2}} , \vspace*{0.3cm}
\end{equation}

\noindent with $\nu$ degrees of freedom, $I(\cdot)$ the indicator function, expectation $\mu$, variance $\sigma^2 = \nu/(\nu-2)$ and $\Gamma(\cdot)$ being the Gamma function. Moreover, $\xi$ is the skewness parameter with $\xi > 0$. It reduces $f$ to the noncentral Student's t-distribution if $\xi = 1$.

Finally, the model description conducts into two approaches, although both models incorporate interaction terms \cite{kreyszig2010advanced}. This idea is based on our data set, where we observe a high periodic structure. Initially, we apply the classical additive Fourier series decomposition, but there is still a periodic behaviour observable. Therefore, we take the periodic interaction into account. The interaction of diurnal periodic functions with the annual ones covers a flexible and time varying explanatory variable within the mean part of \eqref{general}. Depending on the dimension of the indicator matrix, we observe different periodic functions which interact with each other. According to our wind speed models, the Bayesian information criterion (BIC) and the significance of different interactions result in the following indicator matrix

\begin{equation}
\boldsymbol{I} =
\left(\begin{array}{ccccc}
0 & 1 & 1 & 1 & 1 \\
1 & 1 & 1 & 0 & 0 \\
1 & 1 & 1 & 0 & 0 \\
1 & 0 & 0 & 0 & 0 \\
1 & 0 & 0 & 0 & 0
\end{array} \right)
= \left( I_{ i_1 i_2 }  \right), \vspace*{0.3cm}
\end{equation}

\noindent where $\boldsymbol{I}$ is a $5 \times 5$ matrix. The first element ($I_{11}$) is set to zero, because it is already covered by the intercept \eqref{equation:trend}. Obviously, our obtained indicator matrix $\boldsymbol{I}$ excludes a few interactions. These are combinations of half-daily and half-yearly, half-yearly and diurnal as well as half-daily and annual periodic functions. Thus, we obtain the following seasonal term from Equation \eqref{equation:intandseason}

\begin{eqnarray*}
Seasonal_{t} &=& \vartheta_{12} f_{12}^{s_1}(t)f_{12}^{s_2}(t) + \vartheta_{13} f_{13}^{s_1}(t)f_{13}^{s_2}(t)
                + \vartheta_{14} f_{14}^{s_1}(t)f_{14}^{s_2}(t) + \vartheta_{15} f_{15}^{s_1}(t)f_{15}^{s_2}(t) \\
             && + \vartheta_{21} f_{21}^{s_1}(t)f_{21}^{s_2}(t) + \vartheta_{31} f_{31}^{s_1}(t)f_{31}^{s_2}(t)
                + \vartheta_{41} f_{41}^{s_1}(t)f_{41}^{s_2}(t) + \vartheta_{51} f_{51}^{s_1}(t)f_{51}^{s_2}(t) \\
             &&	+\vartheta_{22} f_{22}^{s_1}(t)f_{22}^{s_2}(t) + \vartheta_{32} f_{32}^{s_1}(t)f_{32}^{s_2}(t)
				+\vartheta_{33} f_{33}^{s_1}(t)f_{33}^{s_2}(t)  + \vartheta_{23} f_{23}^{s_1}(t)f_{23}^{s_2}(t) \\
 &=& \vartheta_{12} \cos_{p}\left(\frac{2 \pi t }{s_2 } \right) + \vartheta_{13}  \sin_{p}\left(\frac{2 \pi t }{s_2 } \right)
   + \vartheta_{14} \cos_{p}\left(\frac{4 \pi t }{s_2 } \right) + \vartheta_{15} \sin_{p}\left(\frac{4 \pi t }{s_2 } \right) \\
 &&+ \vartheta_{21} \cos_{p}\left(\frac{2 \pi t }{s_1} \right) + \vartheta_{31}\sin_{p}\left(\frac{2 \pi t }{s_1} \right)
 + \vartheta_{41} \cos_{p}\left(\frac{4 \pi t }{s_1} \right) + \vartheta_{51}\sin_{p}\left(\frac{4 \pi t }{s_1} \right) \\
 && + \vartheta_{22}\cos_{p}\left(\frac{2 \pi t }{s_1} \right) \cos_{p}\left(\frac{2 \pi t }{s_2} \right)
    + \vartheta_{23}\cos_{p}\left(\frac{2 \pi t }{s_1} \right) \sin_{p}\left(\frac{2 \pi t }{s_2} \right) \\
 && + \vartheta_{32}\sin_{p}\left(\frac{2 \pi t }{s_1} \right) \cos_{p}\left(\frac{2 \pi t }{s_2} \right)
    + \vartheta_{33}\sin_{p}\left(\frac{2 \pi t }{s_1} \right) \sin_{p}\left(\frac{2 \pi t }{s_2} \right) \, . \vspace*{0.3cm}
\end{eqnarray*}

\noindent

\subsection{Parameter estimation and in-sample results}

Subsequently, we discuss the parameter estimation of both modelling approaches. The first model contains classical sine and cosine functions, which is described by \eqref{equation:trend}, \eqref{equation:intandseason} and \eqref{equation:arfimaaparch}. Our first model is estimated by a quasi maximum likelihood method, whereas the p-generalisation parameter of our trigonometric functions remains constant (i.e., $p =2$). Unfortunately, the estimation procedure of the second model is more complex. Here, we have to solve nonlinear equations, which is related to the p-generalised periodic function (i.e. $p>0$), to obtain suitable results. Therefore, we primarily estimate the intercept, trend and the seasonal coefficients as well as the parameter $p>0$ for the generalised sine and cosine functions by generalised least squares estimation. These values are only used as starting values. In the next step we estimate the entire second model by quasi maximum likelihood. According to the complete parameter space which might be included into our model, we have to select an information criterion which punishes over-parametrisation. We choose the Bayesian information criterion (BIC) to find the best model order as it is more conservative according to the parameter selection than, e.g., the Akaike information criterion (AIC).

As previously stated, the second wind speed prediction approach uses p-generalised sine and cosine functions ($p> 0$), which are shown in \eqref{equation:funct} and \eqref{equation:funct1}. For each periodic function $f_{i}^{s}$, with $i \in \{i_1,i_2\}$, $i_1 = \{1,...,5\}$, $i_2 = \{1,...,5\}$ and $\boldsymbol{s} \in \{ s_1, s_2 \}$, we include a $\sin_p$ and $\cos_p$ function in the second model. Thus, we have to estimate the parameter set $p \in \{p_{\boldsymbol{s},\sin}, p_{\boldsymbol{s},\cos}\}$ with $p >0$ for each p-generalised trigonometrical function. Bearing in mind that a special case of this function is $p_{\boldsymbol{s},\sin} =  p_{\boldsymbol{s},\cos} = 2$ and we do not estimate the parameter set $p$ for our first model.

Tables \ref{table:parameterestimation:M1} and \ref{table:parameterestimation:M2} provide the estimated parameters for each station. We show the parameter estimates for the model order which is obtained by the smallest BIC as well as the smallest in-sample mean square error. To evaluate the significance of a coefficient, we use a level of significance of $\alpha = 0.01$. We have to test whether the p-generalisation coefficient $p_{\boldsymbol{s},\sin} \neq 2$ and $p_{\boldsymbol{s},\cos} \neq 2$. Indeed, the p-coefficients  are $p_{\boldsymbol{s},\sin} \neq 2$ and $p_{\boldsymbol{s},\cos} \neq 2$. The estimated coefficients for all stations and both modelling approaches are significant, with few exceptions. Apparently, the diurnal and annual seasons are significant as well as the biannual and half-daily ones. Surprisingly, the fractional integration parameter $d$ is rejected at station Lindenberg for both models. According to the long-memory parameter $d \in (-0.5,0.5)$, we obtain different, but significant results for the other stations. The optimal model order for the ARFIMA(j,d,q) part in both cases is  given by $j=2$ autoregressive and $q =1$ moving average parameters. The estimated APARCH coefficients are significant for $\alpha = 0.01$, except for the APARCH intercept $\alpha_0$ for Manschnow and Gr\"unow. Here, the best model order of the APARCH process is given by $Q=2$ and $P=1$. Moreover, $\gamma$ is significant for all stations and models. According to the definition shown in \eqref{equation:arfimaaparch}, we observe negative estimates, but in fact positive shocks. This means that choppy wind increases the variance of the wind speed. The estimation results of the skew-t distribution indicate a left skewed distribution $\xi > 1$ with degrees of freedom $\nu > 2$.  While performing a partial t-test whether $\xi = 1$ or alternatively $\xi \neq 1$, we have to reject the null-hypothesis for each station. Analysing $\nu$ for all examples provides computations of $7 < \nu < 9.5$. This result confirms the heavy tail assumption of the t-distribution.

\begin{center}
    [Table \ref{table:parameterestimation:M1} and \ref{table:parameterestimation:M2} about here.]
\end{center}

\begin{figure}
\begin{center}
\includegraphics[width=15cm, height=4.5cm]{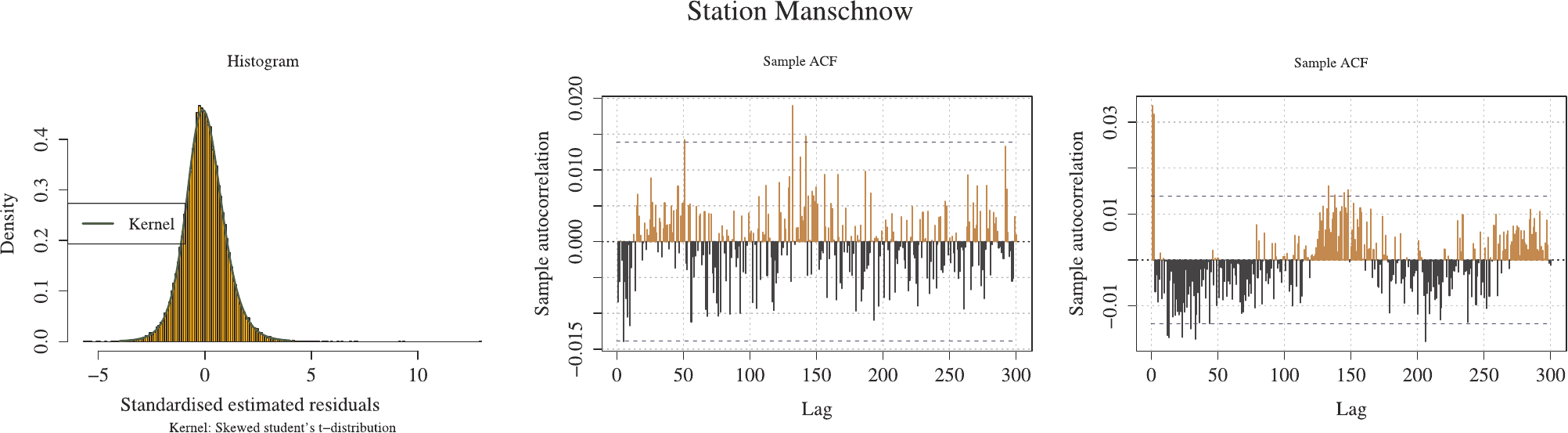} \\
\includegraphics[width=15cm, height=4.5cm]{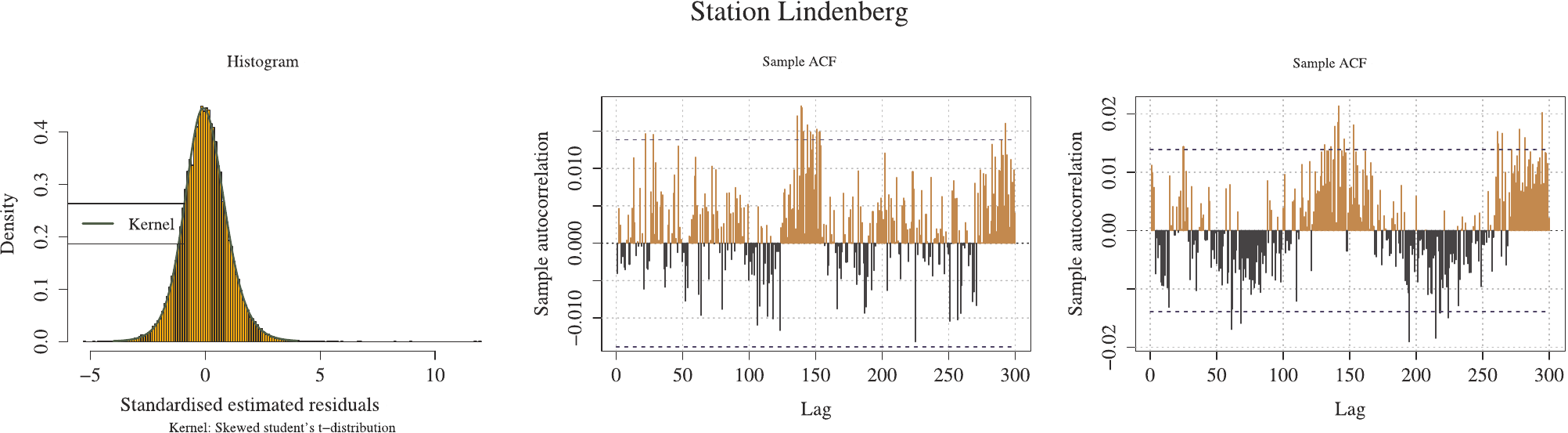} \\
\includegraphics[width=15cm, height=4.5cm]{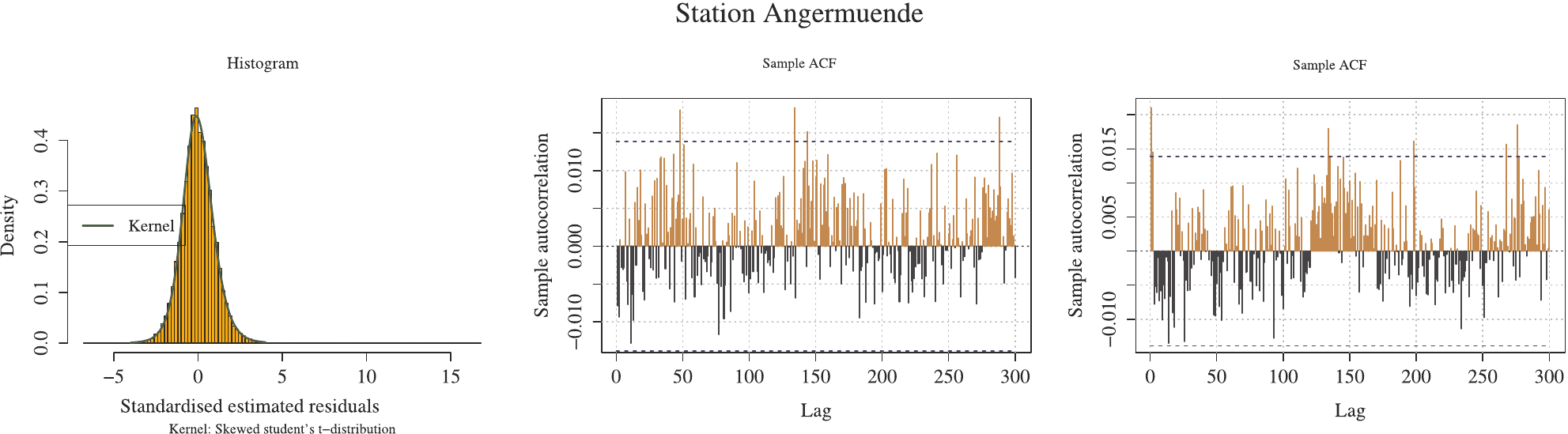} \\
\includegraphics[width=15cm, height=4.5cm]{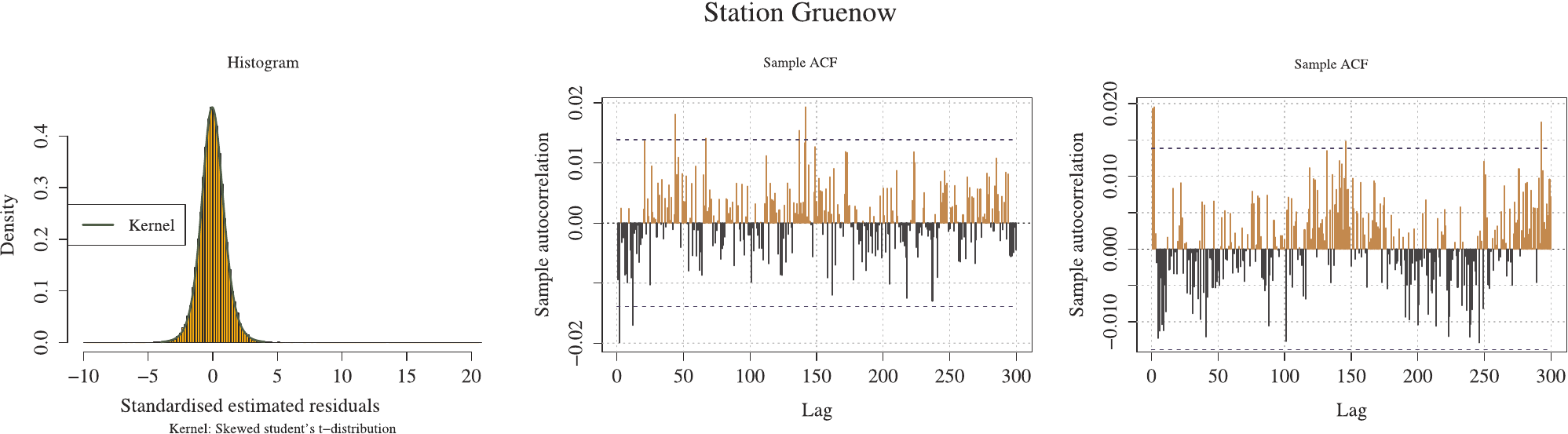} \\
\caption{Model 1: Histogram and ACF plot of $\{\eta\}$ and $\{|\eta|^{\delta}\}$.}
\label{fig:ACF}
\end{center}
The figures show the histogram of the residuals $\{\widehat{\eta_t}\}$ in the first column. The ACF of the residuals are provided in the second column, whereas the third column shows the ACF of $\{|\widehat{\eta_t}|^{\hat{\delta}}\}$.
\end{figure}

\begin{figure}
\begin{center}
\includegraphics[width=15cm, height=4.5cm]{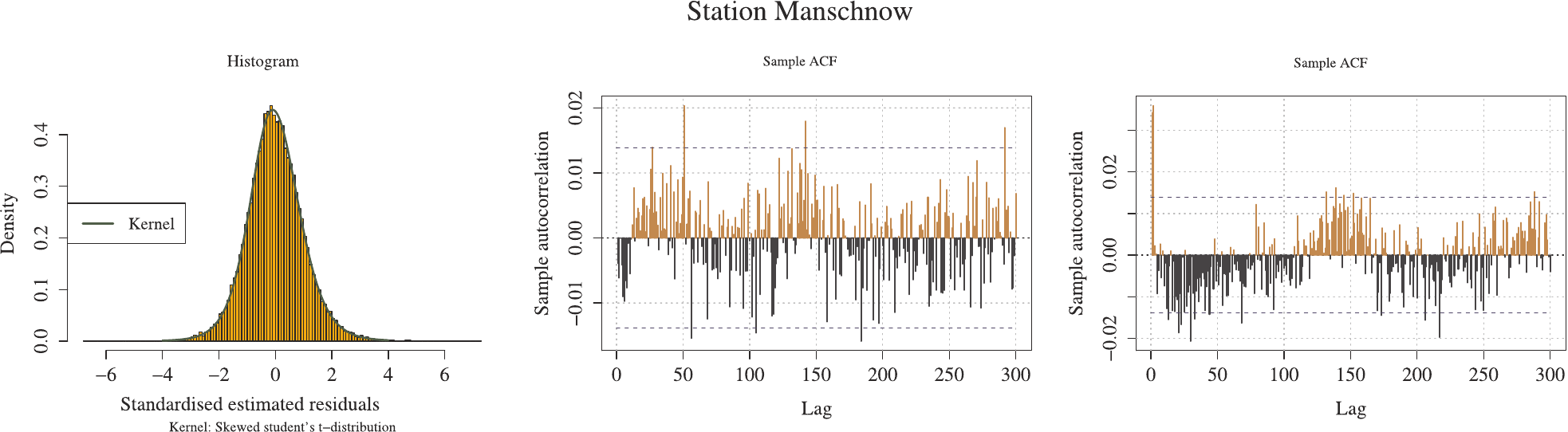} \\
\includegraphics[width=15cm, height=4.5cm]{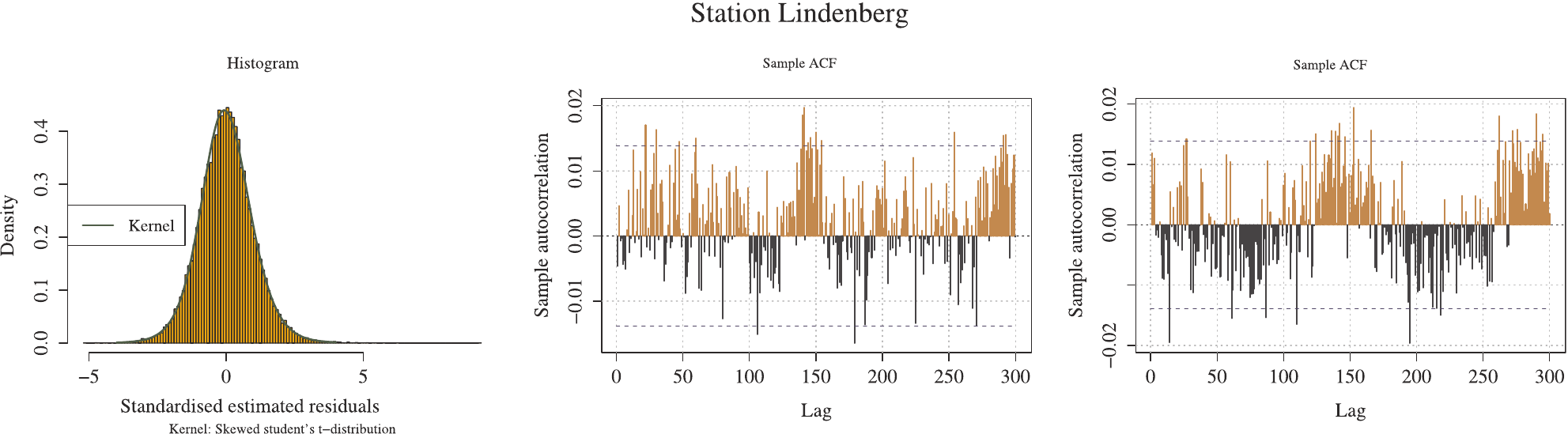} \\
\includegraphics[width=15cm, height=4.5cm]{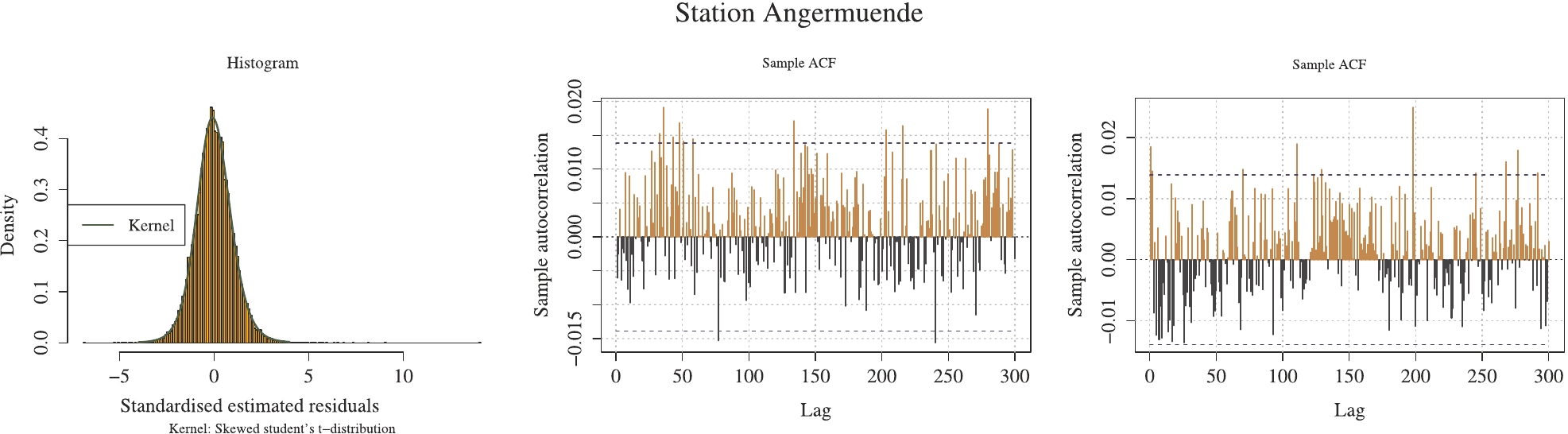} \\
\includegraphics[width=15cm, height=4.5cm]{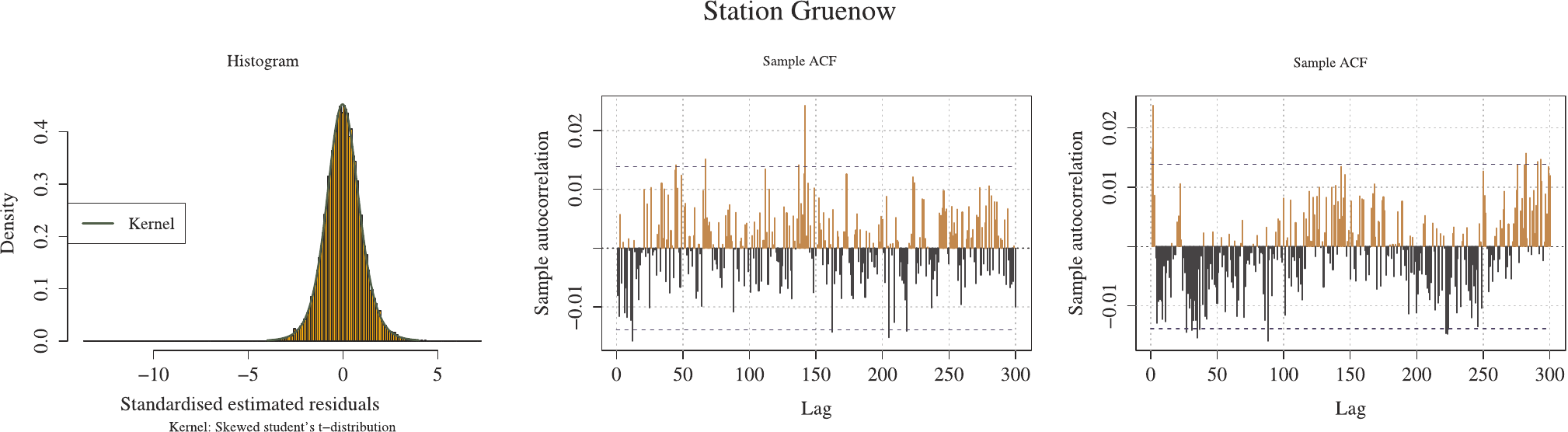} \\
\caption{Histogram, ACF plot of $\eta$ and $|\eta|^{\delta}$ -- second model}
\label{fig:ACF2}
\end{center}
The figures show the histogram of the residuals $\{\widehat{\eta_t}\}$ in the first column. The ACF of the residuals are provided in the second column, whereas the third column shows the ACF of $\{|\widehat{\eta_t}|^{\hat{\delta}}\}$.
\end{figure}

\noindent Following the discussion of the parameter estimation, we analyse the autocorrelation function of the standardised residuals and the absolute value of the residuals with power $\delta$. The results are illustrated in Figure \ref{fig:ACF} and \ref{fig:ACF2}. The histograms of the standardized residuals are shown and compared to the underlying skew-t distribution with skewness parameter $\xi$ and $\nu$ degrees of freedom.  Figure \ref{fig:ACF} provides satisfying results for the standardized residuals $\{\eta_{t}\}$, but with some limitations. One problem is obviously disclosed by the autocorrelation function (ACF) plot. The autocorrelation of the standardised residuals still show  minor periodicity. In comparison to the absolute value of the lagged $\delta$ residuals, the seasonality becomes apparent. According to Figure \ref{fig:ACF2}, the results are very similar to Figure \ref{fig:ACF}. The histogram of the standardized residuals $\{\eta_{t}\}$ indicates a variance reduction. Moreover, the autocorrelation function supports this finding. The ACF of $\{\eta_{t}\}$ provides slightly better results than the first model. Nevertheless, the ACF of $\{|\eta_{t}|^{\delta}\}$ shows a remaining seasonality.

\begin{center}
    [Figure \ref{fig:ACF} and \ref{fig:ACF2} about here.]
\end{center}

\noindent Finally, we compare the model diagnostics. The improvements of the second model that is indicated by the parameter estimation are emphasized by a lower BIC. In addition, the $R^2$ and the MSE provide better results for the second model. The second model discloses better results related to a minor presence of autocorrelation. We perform an autocorrelation test on the standardized and the absolute values of the standardized residuals with power $\delta$ for different lags. The Ljung and Box Q-Statistics indicate a small correlation for some lags, given a level of significance of $\alpha = 0.05$. Nevertheless, the remaining autocorrelation should be negligible, whereas conducting a test on the underlying skew-t distribution sustains the distributional assumption. Moreover, the histogram of the standardized residuals and the assumed underlying skew-t distribution point into the same direction. Summing up the model diagnostics, there is a multitude of arguments for the selected time series model, but with some limitations according to the periodic correlation.

%%%%%%%%%%%%%%%%%%%%%%%%%%%%%%%%%%%%%%%%%%%%%%%%%%% SECTION 3 %%%%%%%%%%%%%%%%%%%%%%%%%%%%%%%%%%%%%%%%%%%%%%%%%%%%%%%%%%

\section{Forecasting wind power}

The two aforementioned wind speed models are evaluated according to their out-of-sample performance. Therefore, we have to examine different accuracy measures. One of those criteria is the root mean square error (RMSE) and another one which we consider is the mean absolute error (MAE). In contrast to the RMSE and MAE, which evaluate the accuracy of the wind speed prediction, we analyse an alternative loss function. This loss function is becoming more common \cite{zhu2012short}.

\begin{figure}
\begin{center}
\includegraphics[width=15cm, height=9cm]{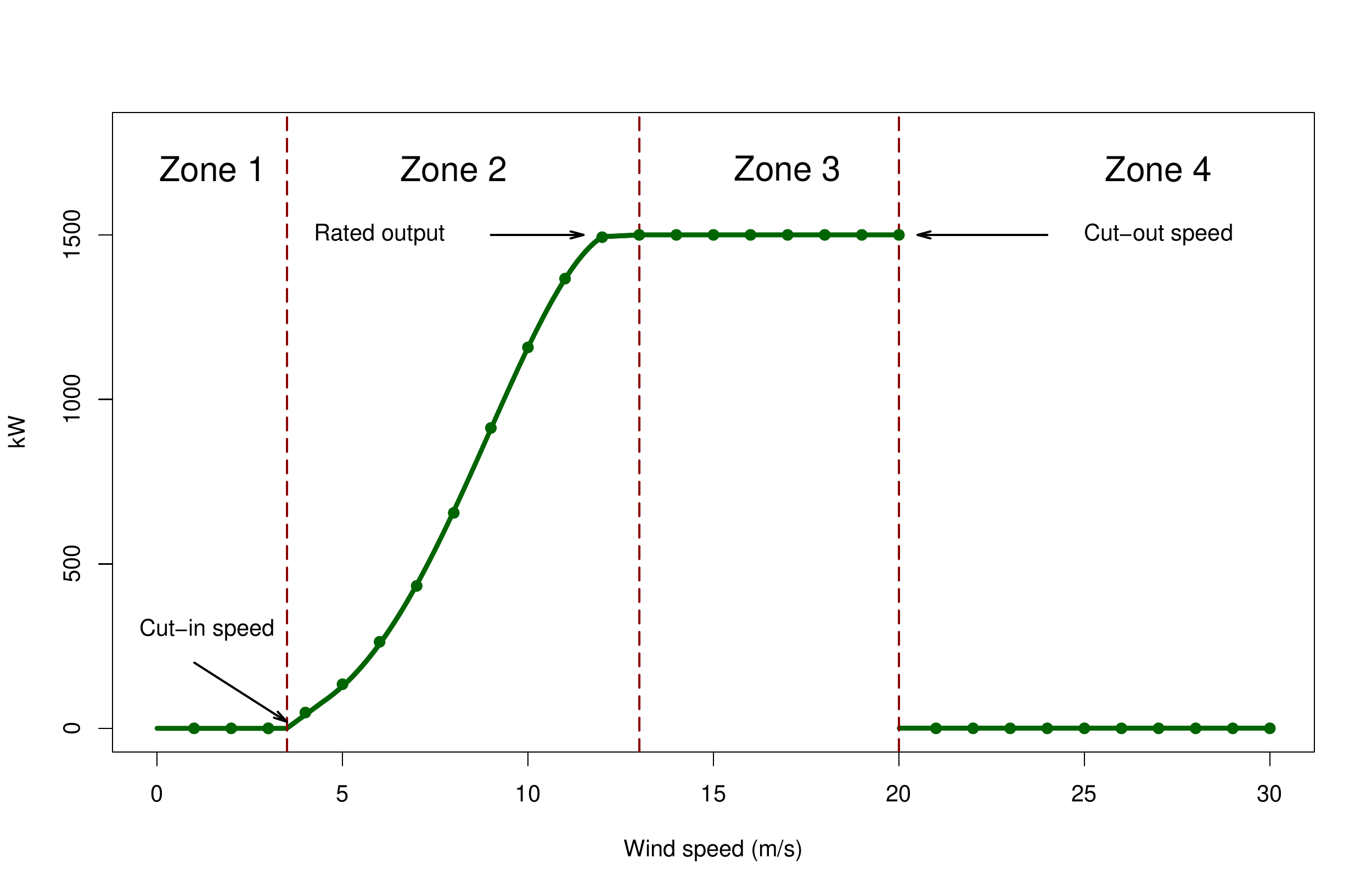} \\
\caption{Power production function of a Fuhrl\"{a}nder MD 77}
\label{fig:power}
\end{center}
The first zone reaches from $0m/s$ to the cut-in wind speed of $3 m/s$. No power is produced within this area. The sigmoidal relationship of the power function \eqref{powerfunction} is used for the second zone. The rated output wind speed of $13 m/s$ and the rated output power of about $1500 kW$ is the beginning of the third zone. From $13 m/s$ to $20 m/s$, the wind power plant is able to produce the maximum wind power. The rotor stops to produce power within the forth zone, beginning from the cut-out speed of $20 m/s$.
\end{figure}

The main interest is the prediction of wind power \cite{madsen2005standardizing}. The production of wind power strongly depends on factors like: air density $\rho$ $(1.25 kg/m^3)$, the rotor swept area $A$, the wind speed $W_t$ and an efficiency or power constant $C_p$ \cite[see][]{burton2001wind}. The power production formula that models the relationship between wind speed and wind power is given by:

\begin{equation}
\label{powerfunction}
Pow_t = \frac{1}{2} C_p  \rho  A (W_t)^3, \vspace*{0.3cm}
\end{equation}

\noindent where $Pow_t$ is the power production function at time $t$. Obviously, for each turbine type we obtain different power curves and a somehow different loss function. Therefore, assuming a typical power curve which holds for each region and all wind turbines is not suitable. At the German-Polish border region for instance, the Fuhrl\"{a}nder MD 77 wind power plant is widely used. Considering Brandenburg altogether, we assess several different turbine types. They differ significantly with respect to their low cut-in wind speed, efficiency, rated output speed and cut-out speed. \cite{Hering2010} introduce a specific accuracy measure, the power curve error (PCE). Their aim is to point out the problem of wind power prediction. We use this power curve error with a specific power function shown in Figure \ref{fig:power}.

\begin{center}
    [Figure \ref{fig:power} about here.]
\end{center}

\noindent A common problem of the wind power prediction is that precise output is not available at each measurement station. Moreover, if wind power data is available, we only observe noisy observations. Therefore, we use \eqref{powerfunction} to approximate the real power data. The PCE loss function of \cite{Hering2010} is given by

\begin{equation}
 L(W_{t+1},\widehat{W}_{t+1}) =
 \begin{cases}
 \begin{array}{ll}
 \tau(Pow(W_{t+1})- Pow(\widehat{W}_{t+1})),     &\qquad \widehat{W}_{t+1} \leq  W_{t+1} \\
 (1-\tau)(Pow(\widehat{W}_{t+1}) -Pow(W_{t+1})), &\qquad \widehat{W}_{t+1} > W_{t+1},
 \end{array}
 \end{cases} \vspace*{0.3cm}
\end{equation}

\noindent where $\tau$ is a weight with $\tau \in [0;1]$. In contrast to \cite{Hering2010}, we consider cases where underestimation is penalized stronger than overestimation, $\tau < 0.5$ and the reverse case $\tau > 0.5$. The main aim of the introduced models is to minimize the respective loss function. According to the PCE, we want to find the quantile $\tau$ minimizing the loss function.

For the out-of-sample forecasts, we select incidentally different points in time $\kappa$ in the out-of-sample period $\kappa \in \{00:00\text{ Januar }1\,,2012,...,23:50\text{ December }31\,,2012\}$. The previously calculated in-sample model order is used as initial estimate. A part of the information set available at time $\kappa$, namely $W_{\kappa-220,000+1},...,W_{\kappa}$ is used for the out-of-sample forecasts.
We calculate $\widehat{W}_{\kappa+18|\kappa}$ and repeat this procedure 5,000 times to obtain suitable forecasting results up to a maximum of three hours. For each model type we use the so called persistence method \cite{nielsen1998new} as a crude benchmark. Such a short-term forecast model uses the observation today as predictor for the next instance in time. The forecast for such a model is quite simple, but provides practicable results for high-frequency data. For the PCE, we take different values for $\tau$. \cite{Hering2010} follow \cite{pinson2007trading}, who propose $\tau = 0.73$. In this case, underestimation is punished more strongly. Moreover, they take different values of $\tau$ to control the overall estimation results and analyse the sensitivity of their models. According to their findings, we decide to choose three different values of $\tau \in \{0.25,0.5,0.75\}$.

Table \ref{table:pred1} gives an overview of predictability results of the three models. Primarily, we discuss the RMSE and MAE and subsequently consider the PCE. Here, the best prediction model is obvious. In this case, the second model with generalised sine and cosine functions and ARFIMA-APARCH process is the best one for a three hour ahead prediction. We have to remark that at station Lindenberg, the fractional integrated part is insignificant. Even though the persistent model is hard to beat for predicting very short periods of time, the model fails to forecast up to three hours.

\begin{center}
    [Table \ref{table:pred1} about here.]
\end{center}

\noindent Apparently, Table \ref{table:pred1} provides an ambiguous overall result.  Analysing the PCE changes our previous findings. As a result from the PCE comparison, we select the first model with ARFIMA-APARCH process and Fourier series in the mean part, if $\tau = 0.25$. The first model has still some minor model limitations related to the remaining information within the variance of the standardised residuals. The forecasting accuracy of model two is inferior to the first model. Still, it is better than the persistent model. This finding indicates that we have to choose the first model if overestimation of wind power is punished. This might be the case if the profit of a wind farm is calculated.

In general, it is important for an investment into a wind farm to know how much wind on average is expected and how much wind power has to be produced. The investor is unwilling to receive losses according to an overestimation of the expected wind power. The PCE for $\tau=0.5$ indicates that the second model provides the best forecasting results. Here, the over- and underestimation is punished similarly. Therefore, it is not surprising that after our transformation into wind power the second model remains the best one. The third row shows the PCE for $\tau = 0.75$. Though the underestimation is penalized, the overall PCE that is shown in the last column provides only marginal differences between each model. According to our previous findings, the best prediction model changes from station to station and also from month to month. The initial best prediction model (Model 2) tends to overestimate the predicted wind power of the Fuhrl\"{a}nder MD 77. Punishing underestimation of wind power is related to the loss function of a system operator. In case of underestimation, the system operator has to order too much power to match the demand. Subsequently, the operator has more power than needed, the generation is reduced and he faces costs of down-regulation. \cite{pinson2007trading} point out that the costs of down-regulation are higher than those for up-regulation and lead to higher expenditures.

Indeed, the results of the PCE are related to the underlying production function shown in Figure \ref{fig:power}. The Fuhrl\"{a}nder MD 77 is a small turbine which is used onshore. We are able to include various turbines to control the sensitivity of the obtained results. While considering different types of turbines, the PCE changes as well. Nevertheless, the turbines onshore situated at the German-Polish border are somehow comparable. Most of them have a similar cut-in wind speed, but a few of them have a higher rated and cut-out wind speed. While changing the wind power production function, the PCE changes in some cases. We use the power production function of a GE 1.6 MW turbine to verify our previous findings. Although some results remain similar, we obtain different findings when $\tau = 0.75$. Now, the second model is marginally better than the other prediction models for forecasting wind power.

The second model provides the best prediction results for the German-Polish border region. Moreover, it is related to the weight $\tau$ which model is the best method to make predictions of the wind power. In particular, the persistent model is a good benchmark for wind power prediction, but for our wind speed prediction, it is outperformed. An onshore wind park operator who is interested in a lower bound of wind power prediction has to use the second model.

%%%%%%%%%%%%%%%%%%%%%%%%%%%%%%%%%%%%%%%%%%%%%%%%%%% SECTION5 %%%%%%%%%%%%%%%%%%%%%%%%%%%%%%%%%%%%%%%%%%%%%%%%%%%%%%%%%%

\section{Conclusion}

Recent research has covered a wide range of wind speed forecasting approaches. In this paper, two novel statistical approaches for high-frequency wind speed data are described. Several extensions are implemented to cover the specific structure of the data set. The mean part takes annual and diurnal interactions into account. Moreover, we distinguish between a first model with classical Fourier series and a second model with p-generalised sine and cosine functions. The autocorrelation and the heteroscedasticity of both approaches is modeled by an ARFIMA-APARCH process with skew-t distributed residuals. The models are suitable to make three hour forecasts for wind speed and can be used for wind power prediction as well.

According to RMSE and MAE, the developed approaches outperform common models. The generalised trigonometrical functions provide the smallest accuracy measures according to the three hour ahead forecasts. Furthermore, we comprise the PCE which provides ambiguous results. The overall PCE provides the best results for the first model, if we consider a small weight.  Nevertheless, the second model derives the best results if the weight is $0.5$ or $0.75$. Finally, we remark that the PCE is also depending on the previously specified turbine type. In this article, the models are applied to four stations situated at the German-Polish border. For this reason, the power production function of a Fuhrl\"{a}nder MD 77 is used to calculate the PCE. The Fuhrl\"{a}nder MD 77 is a small turbine which is often used onshore, especially in plain regions in Germany. The model robustness which is briefly discussed in the end, shows that some changes within the power curve function lead to different best prediction models.

In practice, only the wind speed data set is needed. Therefore, such a statistical wind speed prediction model is cost saving compared to other modelling approaches that require a set of possibly many more variables. Clearly, there are further improvements of this approach possible. An idea is to analyse the variance in more detail and to consider meteorological regressors. Finally, the wind speed model might include spatial informations.

\bibliographystyle{plain}

\begin{thebibliography}{10}

\bibitem{Ambach2015}
Daniel Ambach and Wolfgang Schmid.
\newblock Periodic and long range dependent models for high frequency wind
  speed data.
\newblock {\em Energy}, 82:277 -- 293, 2015.

\bibitem{benth2010analysis}
J{\=u}rat{\.e}~{\v{S}}altyt{\.e} Benth and Fred~Espen Benth.
\newblock Analysis and modelling of wind speed in new york.
\newblock {\em Journal of Applied Statistics}, 37(6):893--909, 2010.

\bibitem{Brockwell2009}
Peter~J Brockwell and Richard~A Davis.
\newblock {\em Time series: theory and methods}.
\newblock Springer, New York, 2009.

\bibitem{burton2001wind}
Tony Burton, David Sharpe, Nick Jenkins, and Ervin Bossanyi.
\newblock {\em Wind energy handbook}.
\newblock John Wiley \& Sons, 2001.

\bibitem{Ding1993}
Zhuanxin Ding, Clive~WJ Granger, and Robert~F Engle.
\newblock A long memory property of stock market returns and a new model.
\newblock {\em Journal of empirical finance}, 1(1):83--106, 1993.

\bibitem{Ewing2006}
Bradley~T. Ewing, Jamie~Brown Kruse, and John~L. Schroeder.
\newblock Time series analysis of wind speed with time-varying turbulence.
\newblock {\em Environmetrics}, 17(2):119--127, Mar 2006.

\bibitem{Fernandez1998}
Carmen Fern{\'a}ndez and Mark~FJ Steel.
\newblock On bayesian modeling of fat tails and skewness.
\newblock {\em Journal of the American Statistical Association},
  93(441):359--371, 1998.

\bibitem{gneiting2006calibrated}
Tilmann Gneiting, Kristin Larson, Kenneth Westrick, Marc~G Genton, and Eric
  Aldrich.
\newblock Calibrated probabilistic forecasting at the stateline wind energy
  center: The regime-switching space--time method.
\newblock {\em Journal of the American Statistical Association},
  101(475):968--979, 2006.

\bibitem{Haslett1989}
John Haslett and Adrian~E Raftery.
\newblock Space-time modelling with long-memory dependence: Assessing ireland's
  wind power resource.
\newblock {\em Applied Statistics}, 30(1):1--50, 1989.

\bibitem{Hering2010}
Amanda~S. Hering and Marc~G. Genton.
\newblock Powering up with space-time wind forecasting.
\newblock {\em Journal of the American Statistical Association},
  105(489):92--104, Mar 2010.

\bibitem{jiang2013very}
Yu~Jiang, Zhe Song, and Andrew Kusiak.
\newblock Very short-term wind speed forecasting with bayesian structural break
  model.
\newblock {\em Renewable energy}, 50:637--647, 2013.

\bibitem{jung2014current}
Jaesung Jung and Robert~P Broadwater.
\newblock Current status and future advances for wind speed and power
  forecasting.
\newblock {\em Renewable and Sustainable Energy Reviews}, 31:762--777, 2014.

\bibitem{kreyszig2010advanced}
Erwin Kreyszig.
\newblock {\em Advanced engineering mathematics}.
\newblock John Wiley \& Sons, 2010.

\bibitem{madsen2005standardizing}
Henrik Madsen, Pierre Pinson, George Kariniotakis, Henrik~Aa Nielsen, and
  Torben~S Nielsen.
\newblock Standardizing the performance evaluation of shortterm wind power
  prediction models.
\newblock {\em Wind Engineering}, 29(6):475--489, 2005.

\bibitem{nielsen1998new}
Torben~Skov Nielsen, Alfred Joensen, Henrik Madsen, Lars Landberg, and Gregor
  Giebel.
\newblock A new reference for wind power forecasting.
\newblock {\em Wind energy}, 1(1):29--34, 1998.

\bibitem{pinson2007trading}
Pierre Pinson, Christophe Chevallier, and George~N Kariniotakis.
\newblock Trading wind generation from short-term probabilistic forecasts of
  wind power.
\newblock {\em Power Systems, IEEE Transactions on}, 22(3):1148--1156, 2007.

\bibitem{richter2007generalized}
Wolf-Dieter Richter.
\newblock Generalized spherical and simplicial coordinates.
\newblock {\em Journal of Mathematical Analysis and Applications},
  336(2):1187--1202, 2007.

\bibitem{richter2011ellipses}
Wolf-Dieter Richter.
\newblock Ellipses numbers and geometric measure representations.
\newblock {\em Journal of Applied Analysis}, 17(2):165--179, 2011.

\bibitem{soman2010review}
Saurabh~S Soman, Hamidreza Zareipour, Om~Malik, and Paras Mandal.
\newblock A review of wind power and wind speed forecasting methods with
  different time horizons.
\newblock In {\em North American Power Symposium (NAPS), 2010}, pages 1--8.
  IEEE, 2010.

\bibitem{Taylor2009}
James~W Taylor, Patrick~E McSharry, and Roberto Buizza.
\newblock Wind power density forecasting using ensemble predictions and time
  series models.
\newblock {\em Energy Conversion, IEEE Transactions on}, 24(3):775--782, 2009.

\bibitem{europeanwind}
E.~W. E.~A. The European Wind Energy~Association.
\newblock Wind in power 2013 european statistics february 2014.
\newblock {\em EWEA: Brussels}, 2014.

\bibitem{WWEA2014}
W.~W. E.~A. The World Wind Energy~Association.
\newblock World wind energy half-year report 2014.
\newblock {\em WWEA: Bonn}, 2014.

\bibitem{zhu2012short}
Xinxin Zhu and Marc~G Genton.
\newblock Short-term wind speed forecasting for power system operations.
\newblock {\em International Statistical Review}, 80(1):2--23, 2012.

\bibitem{zhu2014space}
Xinxin Zhu, Marc~G Genton, Yingzhong Gu, and Le~Xie.
\newblock Space-time wind speed forecasting for improved power system dispatch.
\newblock {\em Test}, 23(1):1--25, 2014.

\end{thebibliography}

\newpage

\newpage

\newpage

\begin{table}
\begin{tiny}
\caption{Parameter estimation of the first model for all stations.}
\begin{center}
    \begin{tabular}{rrrrrrrrr}    \hline
  &  \multicolumn{2}{c}{\textbf{Manschnow}}  & \multicolumn{2}{c}{\textbf{Lindenberg}}   & \multicolumn{2}{c}{\textbf{Angerm\"unde}} & \multicolumn{2}{c}{\textbf{Gr\"unow}}  \\
\hline
     \multicolumn{9}{l}{\textbf{Regression coefficients}}  \\
          & Estimate & Std. Error &   Estimate & Std. Error  & Estimate & Std. Error &   Estimate & Std. Error  \\
$\vartheta_{trend}$& -0.0000025$^{***}$ & 0.00000006&-0.0000040$^{***}$&0.00000006&-0.0000027$^{***}$&0.00000006&-0.0000028$^{***}$& 0.00000006  \\
  $\vartheta_{11}$ &    3.2941$^{***}$  & 0.0075    &  4.0346$^{***}$  & 0.0079   &  4.1071$^{***}$  & 0.0089   &  4.6358$^{***}$  & 0.0093      \\
  $\vartheta_{12}$ &    -0.5588$^{***}$ & 0.0053    &  -0.1456$^{***}$ & 0.0056   &  -0.5420$^{***}$ & 0.0062   &  -0.5464$^{***}$ & 0.0065      \\
  $\vartheta_{13}$ &    -0.3730$^{***}$ & 0.0053    &  -0.1107$^{***}$ & 0.0056   &  -0.4179$^{***}$ & 0.0062   &  -0.4529$^{***}$ & 0.0065      \\
  $\vartheta_{14}$ &    0.0752$^{***}$  & 0.0053    &  0.0313$^{***}$  & 0.0056   &  0.0743$^{***}$  & 0.0062   &  0.1005$^{***}$  & 0.0065      \\
  $\vartheta_{15}$ &    0.2098$^{***}$  & 0.0053    &  0.1089$^{***}$  & 0.0056   &  0.2144$^{***}$  & 0.0062   &  0.1938$^{***}$  & 0.0065      \\
  $\vartheta_{21}$ &    0.0451$^{***}$  & 0.0053    &  -0.2839$^{***}$ & 0.0056   &  -0.3678$^{***}$ & 0.0062   &  -0.1986$^{***}$ & 0.0065      \\
  $\vartheta_{22}$ &    0.0531$^{***}$  & 0.0074    &  0.0057$^{***}$  & 0.0078   &  -0.0888$^{***}$ & 0.0088   &  -0.1646$^{***}$ & 0.0092      \\
  $\vartheta_{23}$ &    0.1079$^{***}$  & 0.0074    &  -0.0334$^{***}$ & 0.0078   &  0.0831$^{***}$  & 0.0088   &  -0.0139$^{***}$ & 0.0092      \\
  $\vartheta_{31}$ &    0.6339$^{***}$  & 0.0053    &  0.4167$^{***}$  & 0.0056   &  0.5400$^{***}$  & 0.0063   &  0.6472$^{***}$  & 0.0066      \\
  $\vartheta_{32}$ &    0.4127$^{***}$  & 0.0075    &  0.0656$^{***}$  & 0.0079   &  0.3795$^{***}$  & 0.0088   &  0.3441$^{***}$  & 0.0093      \\
  $\vartheta_{33}$ &    0.2368$^{***}$  & 0.0075    &  0.1021$^{***}$  & 0.0079   &  0.3114$^{***}$  & 0.0088   &  0.2975$^{***}$  & 0.0093      \\
  $\vartheta_{41}$ &    -0.0628$^{***}$ & 0.0053    &  -0.0418$^{***}$ & 0.0056   &  0.0166$^{***}$  & 0.0062   &  0.1447$^{***}$  & 0.0065      \\
  $\vartheta_{51}$ &    -0.1060$^{***}$ & 0.0053    &  -0.1549$^{***}$ & 0.0056   &  -0.1123$^{***}$ & 0.0062   &  -0.2368$^{***}$ & 0.0065      \\
    \multicolumn{9}{l}{\textbf{ARFIMA parameters}}  \\
    $ d      $ & 0.4310$^{***}$  & 0.0064 &  0.00003      & 0.0003 & 0.0913$^{***}$   &  0.0275 &  0.2323$^{***}$  & 0.0164 \\
    $\phi_1$   & 1.2334$^{***}$  & 0.0033 &  1.4033$^{***}$ & 0.0071 & 1.5345$^{***}$   &  0.0168 &  1.4446$^{***}$  & 0.0035 \\
    $\phi_2$   &-0.2698$^{***}$  & 0.0029 & -0.4104$^{***}$ & 0.0077 &-0.5412$^{***}$   &  0.0162 & -0.4562$^{***}$  & 0.0032 \\
    $\theta_1$ &-0.8065$^{***}$  & 0.0067 & -0.6186$^{***}$ & 0.0102 &-0.7650$^{***}$   &  0.0133 & -0.7826$^{***}$  & 0.0126 \\
    \multicolumn{9}{l}{\textbf{APARCH parameters}}  \\
    $\alpha_0$ &   0.0012$^{**}$ & 0.0004 &  0.0091$^{***}$ & 0.0009 &  0.0188$^{***}$    & 0.0017 &   0.0019$^{*}$   & 0.0012 \\
    $\alpha_1$ &   0.1426$^{***}$& 0.0054 &  0.1346$^{***}$ & 0.0075 &  0.1254$^{***}$    & 0.0071 &   0.1580$^{***}$ & 0.0064\\
    $\beta_1$  &   0.5257$^{***}$& 0.0459 &  0.7160$^{***}$ & 0.0622 &  0.5523$^{***}$    & 0.0533 &   0.5526$^{***}$ & 0.0496\\
    $\beta_2$  &   0.3645$^{***}$& 0.0433 &  0.1618$^{***}$ & 0.0569 &  0.3015$^{***}$    & 0.0490 &   0.3252$^{***}$ & 0.0460\\
    $\gamma_1$ &  -0.1208$^{***}$& 0.0179 & -0.2860$^{***}$ & 0.0238 & -0.1947$^{***}$    & 0.0255 &  -0.1103$^{***}$ & 0.0187\\
    $\delta$   &   0.9325$^{***}$& 0.0606 &  0.9529$^{***}$ & 0.0572 &  1.3877$^{***}$    & 0.0770 &   1.0750$^{***}$ & 0.0543\\
    $\xi$      &   1.0672$^{***}$& 0.0083 &  1.0512$^{***}$ & 0.0083 &  1.0534$^{***}$    & 0.0084 &   1.0160$^{***}$ & 0.0080\\
    $\nu$      &   7.8622$^{***}$& 0.3101 &  9.2202$^{***}$ & 0.4505 &  8.8646$^{***}$    & 0.4003 &   7.3534$^{***}$ & 0.2841\\

    \textbf{BIC} & &   1.2886   &   & 1.1598  & &  1.5658   &  & 1.4116              \\
    \end{tabular}
    \label{table:parameterestimation:M1}
    $^{***}$ Significance at 1$\%$, $^{**}$ significance at 5$\%$ and $^{*}$ significance at 10$\%$.
\end{center}
\end{tiny}
\end{table}

\newpage

\begin{table}
\begin{tiny}
\caption{Parameter estimation of the second model for all stations.}
\begin{center}
    \begin{tabular}{rrrrrrrrr}    \hline
  &  \multicolumn{2}{c}{\textbf{Manschnow}}  & \multicolumn{2}{c}{\textbf{Lindenberg}}   & \multicolumn{2}{c}{\textbf{Angerm\"unde}} &
  \multicolumn{2}{c}{\textbf{Gr\"unow}}  \\
\hline
     \multicolumn{9}{l}{\textbf{Regression coefficients}}  \\
          & Estimate & Std. Error &   Estimate & Std. Error  & Estimate & Std. Error &   Estimate & Std. Error  \\
$\vartheta_{trend}$&-0.0000025$^{***}$&0.00000006&-0.0000040$^{***}$&0.00000006&-0.0000027$^{***}$& 0.00000007&   -0.0000028$^{***}$&0.00000007 \\
$\vartheta_{11}$   &   3.2938$^{***}$ & 0.0075   &  4.0356$^{***}$  & 0.0079   & 4.1064$^{***}$   & 0.0089    & 4.6355$^{***}    $&  0.0093    \\
$\vartheta_{12}$   &   -0.5387$^{***}$& 0.0051   & -0.1407$^{***}$  & 0.0054   & -0.5227$^{***}$  & 0.0060    & -0.5269$^{***} $  &   0.0063    \\
$\vartheta_{13}$   &   -0.3987$^{***}$& 0.0056   & -0.1183$^{***}$  & 0.0060   & -0.4452$^{***}$  & 0.0066    & -0.4839$^{***} $  &   0.0070    \\
$\vartheta_{14}$   &   0.0666$^{***}$ & 0.0047   & 0.0274$^{***}$   & 0.0049   & 0.0658$^{***}$   & 0.0055    &  0.0890$^{***} $  &    0.0058    \\
$\vartheta_{15}$   &   0.2100$^{***}$ & 0.0053   & 0.1090$^{***}$   & 0.0218   & 0.2147$^{***}$   & 0.0062    &  0.1941$^{***} $  &    0.0065    \\
$\vartheta_{21}$   &   0.0410$^{***}$ & 0.0047   & -0.2354  $^{***}$& 0.0049   & -0.3147$^{***}$  & 0.0055    & -0.1672$^{***} $  & 0.0058    \\
$\vartheta_{22}$   &   0.1612$^{***}$ & 0.0207   & 0.0501   $^{**}$ & 0.0218   & -0.2078$^{***}$  & 0.0244    &  -0.4095$^{***}$  &  0.0255    \\
$\vartheta_{23}$   &   7.6052$^{***}$ & 0.4566   & -1.6406  $^{***}$& 0.4824   & 2.7394$^{***}$   & 0.5387    &  -1.4781$^{***}$  & 0.5633    \\
$\vartheta_{31}$   &   0.5649$^{***}$ & 0.0047   & 0.3787   $^{***}$& 0.0050   & 0.4958$^{***}$   & 0.0056    &  0.5853$^{***} $  &  0.0058    \\
$\vartheta_{32}$   &   0.5948$^{***}$ & 0.0107   & 0.0797   $^{***}$& 0.0113   & 0.5288$^{***}$   & 0.0126    &  0.4836$^{***} $  & 0.0132    \\
$\vartheta_{33}$   &   0.4507$^{***}$ & 0.0137   & 0.1920   $^{***}$& 0.0144   & 0.5904$^{***}$   & 0.0161    &  0.5665$^{***} $  &  0.0168    \\
$\vartheta_{41}$   &   -0.1658$^{***}$& 0.0107   & -0.0067          & 0.0113   & 0.1490$^{***}$   & 0.0126    &  0.0890$^{***}$   & 0.0058    \\
$\vartheta_{51}$   &   -0.1030$^{***}$& 0.0050   & -0.1517  $^{***}$& 0.0053   & -0.1093$^{***}$  & 0.0059    &  0.1941$^{***}$   &  0.0065    \\
$p_{12}$           &    2.4709$^{***}$& 0.0104   &  4.5886  $^{***}$& 0.0029   &    1.9080$^{***}$& 0.0102    &   2.3876$^{***}$  &  0.0024    \\
$p_{13}$           &    1.0264$^{***}$& 0.0058   &  1.4824  $^{***}$& 0.0039   &    1.1167$^{***}$& 0.0057    &   1.0529$^{***}$  & 0.0036    \\
$p_{14}$           &    1.3801$^{***}$& 0.0432   &  51.3130 $^{***}$& 0.1569   &    1.2694$^{***}$& 0.0091    &   1.6574$^{***}$  & 0.0026    \\
$p_{15}$           &    1.7776$^{***}$& 0.0189   &  1.5631  $^{***}$& 0.0019   &    1.9647$^{***}$& 0.0104    &   4.2203$^{***}$  & 0.0035    \\
$p_{21}$           &    0.3945$^{***}$& 0.0030   &  0.4286  $^{***}$& 0.0009   &    0.1752$^{***}$& 0.0002    &   0.1566$^{***}$  & 0.0002    \\
$p_{31}$           &    8.5535$^{***}$& 0.1220   & 31.0516  $^{***}$& 0.1011   &   30.9854$^{***}$& 0.0146    &  19.7518$^{***}$  & 0.0047    \\
$p_{41}$           &    0.5056$^{***}$& 0.0060   &  76.5876 $^{***}$& 0.2387   &    0.1535$^{***}$& 0.0004    &   0.1404$^{***}$  &  0.0001    \\
$p_{51}$           &    1.6385$^{***}$& 0.0485   &  15.3874 $^{***}$& 0.0409   &    6.9258$^{***}$& 0.0223    &   0.1778$^{***}$  &  0.0002    \\
    \multicolumn{9}{l}{\textbf{ARFIMA parameters}}  \\
$ d      $ & 0.4475$^{***}$& 0.0024& 0.000001        &0.000008&  0.1006$^{***}$ &   0.0126 &  0.2564$^{***}$  &  0.0087\\
$\phi_1$   & 1.2184$^{***}$& 0.0094&  1.4044 $^{***}$& 0.0011 &  1.4961$^{***}$ &   0.0183 &  1.4318$^{***}$  &  0.0016\\
$\phi_2$   &-0.2583$^{***}$& 0.0071& -0.4116 $^{***}$& 0.0011 & -0.5048$^{***}$ &   0.0172 & -0.4438$^{***}$  &  0.0015\\
$\theta_1$ &-0.8081$^{***}$& 0.0215& -0.6195 $^{***}$& 0.0047 & -0.7683$^{***}$ &   0.0133 & -0.7919$^{***}$  &  0.0014\\
\multicolumn{9}{l}{\textbf{APARCH parameters}}  \\
$\alpha_0$ & 0.0012$^{*}$  &  0.0006 &  1.75 & 0.0091$^{***}$ &  0.0189$^{***}$  &  0.0017  &  0.0020$^{*}$    &  0.0011\\
$\alpha_1$ & 0.1392$^{***}$&  0.0051 & 26.79 & 0.1343$^{***}$ &  0.1260$^{***}$  &  0.0071  &  0.1583$^{***}$  &  0.0064\\
$\beta_1$  & 0.5188$^{***}$&  0.0463 & 11.18 & 0.7175$^{***}$ &  0.5495$^{***}$  &  0.0529  &  0.5516$^{***}$  &  0.0494\\
$\beta_2$  & 0.3738$^{***}$&  0.0437 &  8.54 & 0.1606$^{***}$ &  0.3037$^{***}$  &  0.0486  &  0.3259$^{***}$  &  0.0458\\
$\gamma_1$ &-0.1372$^{***}$&  0.0131 &-10.46 &-0.2863$^{***}$ & -0.1951$^{***}$  &  0.0254  & -0.1086$^{***}$  &  0.0188\\
$\delta$   & 0.9594$^{***}$&  0.1076 &  8.91 & 0.9430$^{***}$ &  1.3879$^{***}$  &  0.0766  &  1.0809$^{***}$  &  0.0463\\
$\xi$      & 1.0693$^{***}$&  0.0083 &128.16 & 1.0517$^{***}$ &  1.0549$^{***}$  &  0.0085  &  1.0158$^{***}$  &  0.0079\\
$\nu$      & 7.8890$^{***}$&  0.2934 & 26.88 & 9.1984$^{***}$ &  8.8649$^{***}$  &  0.4010  &  7.3345$^{***}$  &  0.2829\\
\textbf{BIC} & &   1.2695   &  & 1.1185   &   & 1.5616  &      & 1.3596            \\

    \end{tabular}
    \label{table:parameterestimation:M2}
        $^{***}$ Significance at 1$\%$, $^{**}$ significance at 5$\%$ and $^{*}$ significance at 10$\%$.
\end{center}
\end{tiny}
\end{table}

\newpage

\newpage

\newpage

\newpage

\setcounter{table}{3}
\begin{sidewaystable}
\begin{tiny}
\caption{Forecasting accuracy measures for station Manschnow and Lindenberg. The table provides the estimated forecasting measures for each model and station. The lowest of each measures is bold.}
\begin{center}
    \begin{tabular}{r|rrr rrr rrr rrr|r}    
    \hline 
 Model & January & February & March & April & May & June & July & August & September & October & November & December & Total \\

\hline

 \multicolumn{14}{l}{\textbf{Manschnow}} \\
Model 1 RMSE    & 1.61& 1.67& 1.59& 1.63& 1.61& 1.63& 1.61& 1.53& 1.51& 1.63& 1.58& 1.66& 1.61 \\
Model 2 RMSE    &\textbf{1.61}& \textbf{1.66}& \textbf{1.55}& \textbf{1.58}& \textbf{1.52}& \textbf{1.57}& \textbf{1.59}& \textbf{1.51}& \textbf{1.44}& \textbf{1.62}& \textbf{1.55}& \textbf{1.61}& \textbf{1.57} \\
Persistent RMSE & 2.26& 1.91& 1.67& 1.23& 1.42& 1.44& 1.65& 1.41& 1.36& 1.92& 2.17& 2.21& 1.72 \\
\\
Model 1 MAE     & 1.43& 1.51& 1.43& 1.45& 1.46& 1.45& 1.41& 1.37& 1.33& 1.44& 1.43& 1.51& 1.44\\
Model 2 MAE     & \textbf{1.42}& \textbf{1.47}& \textbf{1.38}&\textbf{ 1.41}& \textbf{1.36}& \textbf{1.41}& \textbf{1.39}& \textbf{1.34}& \textbf{1.27}& \textbf{1.43}& \textbf{1.41}& \textbf{1.45}& \textbf{1.40}\\
Persistent MAE  & 2.15& 1.81& 1.59& 1.12& 1.33& 1.34& 1.55& 1.31& 1.28& 1.82& 2.11& 2.16& 1.63\\
\\
Model 1 PCE $\tau = 0.25$   &\textbf{26.11}&\textbf{22.36}&\textbf{16.13}&\textbf{5.83}&\textbf{8.59}& \textbf{6.45}& \textbf{15.41}& \textbf{12.15}&\textbf{6.92}& \textbf{33.19}& \textbf{27.79}& \textbf{37.15}&\textbf{16.42}\\
Model 2 PCE $\tau = 0.25$   &34.18         & 33.21        & 25.21        & 5.95        & 9.16& 7.12& 23.17& 13.24& 7.81& 37.97& 37.84& 59.15&23.128\\
Persistent PCE $\tau = 0.25$&42.15         & 44.16        & 34.39        & 6.18        & 9.76& 7.81& 31.75& 14.33& 8.69& 42.74& 47.91& 81.14&30.92\\
\\
Model 1 PCE $\tau = 0.5$   & 33.24& 33.18  & 21.47& 6.41& 9.21 & 13.17& 14.13& 12.37& 11.92& 45.91& 38.59& 37.88& 23.12\\
Model 2 PCE $\tau = 0.5$   & \textbf{31.16}& \textbf{23.11}& \textbf{14.27}& \textbf{5.84}& \textbf{5.85}&  \textbf{8.84}&  \textbf{9.51}& \textbf{8.28}&  \textbf{9.59}& \textbf{26.17}& \textbf{25.91}& \textbf{28.55}&\textbf{16.42}\\
Persistent PCE $\tau = 0.5$& 36.32& 43.34  & 26.66& 6.95& 12.58& 17.31& 18.56& 16.45& 14.25& 65.65& 51.28& 47.21& 29.71 \\
\\
Model 1 PCE $\tau = 0.75$   & 14.19         &\textbf{19.86}&\textbf{9.68}&\textbf{4.11}&\textbf{7.38}&\textbf{11.71}& 11.56        & 5.37        &  5.21       & 21.18& 18.95        &\textbf{17.65}& 12.24\\
Model 2 PCE $\tau = 0.75$   & \textbf{11.27}& 21.41        & 12.24       & 6.94        &  9.37       & 11.89        &  \textbf{9.91}        & \textbf{4.95}                          &\textbf{4.21} &\textbf{18.47}&\textbf{14.12}& 19.44&\textbf{12.02}\\
Persistent PCE $\tau = 0.75$& 12.73         & 21.64        & 11.96       & 5.52        & 8.38        & 11.81        &11.23&5.16&4.71 &  19.32        & 16.53   & 18.55& 12.30\\
\\
    \multicolumn{14}{l}{\textbf{Lindenberg}}  \\

Model 1 RMSE                &  1.59 &  1.56 &  1.74 &  1.54 &  1.61 &  1.73 &  1.71 &  1.66 &  1.51 &  1.57 &  1.53 & 1.58 & 1.61 \\
Model 2 RMSE                &  \textbf{1.58}&  \textbf{1.53}&  \textbf{1.71}&  \textbf{1.53}&  \textbf{1.58}&  \textbf{1.72} &  \textbf{1.68} &  \textbf{1.61} &  \textbf{1.48} &  \textbf{1.56} &  \textbf{1.51} & \textbf{1.55} & \textbf{1.59} \\
Persistent RMSE             &  2.28 &  2.17 &  1.97 &  1.63 &  1.59 &  1.79 &  1.75 &  1.75 &  1.61 &  2.17 &  1.87 & 2.15 & 1.89 \\
\\
Model 1 MAE                 &  1.41 &  1.38 &  1.55 &  1.41 &  1.42 &  1.52 &  1.51 &  1.48 &  1.34 &  1.42 &  1.36 & 1.41 & 1.43 \\
M2 MAE                      &  \textbf{1.41}&  \textbf{1.35}&  \textbf{1.51}&  \textbf{1.38}&  \textbf{1.39}&  \textbf{1.49}&  \textbf{1.47} &  \textbf{1.42} &  \textbf{1.32} &  \textbf{1.41} &  \textbf{1.34} & \textbf{1.41} & \textbf{1.41} \\
Persistent MAE              &  2.22 &  2.11 &  1.88 &  1.56 &  1.49 &  1.68 &  1.64 &  1.66 &  1.52 &  2.11 &  1.81 & 2.18 & 1.82 \\
\\
Model 1 PCE $\tau = 0.25$   & \textbf{44.38} & \textbf{37.36} & \textbf{42.47} & \textbf{18.56} &  \textbf{9.68} & \textbf{18.77} & \textbf{18.11} & \textbf{16.13} & \textbf{17.96} & \textbf{16.35} & \textbf{39.82} & \textbf{31.21} & \textbf{25.90} \\
Model 2 PCE $\tau = 0.25$   & 59.64 & 51.42 & 55.58 & 28.18 & 13.42 & 21.71 & 24.96 & 27.18 & 26.91 & 25.69 & 44.76 &38.13 &34.80 \\
Persistent PCE $\tau = 0.25$& 74.91 & 65.47 & 68.69 & 37.59 & 17.16 & 24.63 & 31.93 & 38.14 & 35.83 & 35.14 & 49.71 &46.16 &43.78 \\
\\
Model 1 PCE $\tau = 0.5$    & 63.59 & 61.31 & 69.81 & 21.39 & 15.54 & 37.45 & 28.99 & 24.59 & 25.31 & 38.47 & 51.92 &37.91 &39.69 \\
Model 2 PCE $\tau = 0.5$    & \textbf{51.55}& \textbf{47.39}& \textbf{41.61}& \textbf{14.71}& \textbf{14.99}& \textbf{22.16}& \textbf{21.14} & \textbf{15.98} & \textbf{16.13} & \textbf{25.15} & \textbf{32.78} & \textbf{32.61} & \textbf{28.02} \\
Persistent PCE $\tau = 0.5$ & 76.63 & 73.21 & 97.99 & 28.17 & 16.19 & 52.73 & 36.94 & 33.21 & 34.56 & 51.78 & 71.17 &43.22 &51.32 \\
\\
Model 1 PCE $\tau = 0.75$   &23.57 &15.91 &36.87 &12.17 & 8.44 &21.92 &14.58&17.19 &13.12 &15.35 &19.74 &13.99 &17.74 \\
Model 2 PCE $\tau = 0.75$   &25.39 &16.74 & \textbf{34.88} &13.95 &  \textbf{8.43} &21.76 & \textbf{13.52} & \textbf{15.19} & \textbf{11.43} & \textbf{13.51} &21.93 &15.17 & \textbf{17.66} \\
Persistent PCE $\tau = 0.75$& \textbf{21.74} & \textbf{15.17} &38.87 & \textbf{11.21} & 8.46 & \textbf{21.17} &15.64 &18.98 &14.82 &17.21 & \textbf{18.54} & \textbf{12.91} &17.89 \\

\label{table:pred1}
\end{tabular}
\end{center}
\end{tiny}
\end{sidewaystable}

\newpage

\setcounter{table}{4}
\begin{sidewaystable}
\begin{tiny}
\caption{Forecasting accuracy measures for station Angerm\"unde and Gr\"unow. The Table provides the estimated forecasting measures for each model and station. The lowest of each measures is bold.}
\begin{center}
    \begin{tabular}{r|rrr rrr rrr rrr|r}    
    \hline  
Model & January & February & March & April & May & June & July & August & September & October & November & December & Total \\

\hline

    \multicolumn{14}{l}{\textbf{Angerm\"unde}}  \\
Model 1 RMSE    &  1.74 &  1.69 &  1.92 &  1.77 &  1.75 &  1.78 &  1.81 &  1.79 &  1.72 &  1.66 &  1.76 &  1.64 &  1.75 \\
Model 2 RMSE    &  \textbf{1.69} &  \textbf{1.66} &  \textbf{1.86} &  \textbf{1.71} &  \textbf{1.71} &  \textbf{1.73} &  \textbf{1.72} &  \textbf{1.76} &  \textbf{1.65} &  \textbf{1.65} &  \textbf{1.71} &  \textbf{1.61} &  \textbf{1.71} \\
Persistent RMSE &  2.15 &  2.34 &  2.19 &  1.81 &  1.98 &  2.15 &  1.88 &  1.82 &  1.88 &  2.41 &  2.41 &  2.61 &  2.14 \\
\\
Model 1 MAE     &  1.54 &  1.49 &  1.61 &  1.56 &  1.55 &  1.53 &  1.59 &  1.55 &  1.52 &  1.47 &  1.56 &  1.43 &  1.53 \\
Model 2 MAE     &  \textbf{1.48} &  \textbf{1.45} &  \textbf{1.55} &  \textbf{1.49} &  \textbf{1.49} &  \textbf{1.49} &  \textbf{1.46} &  \textbf{1.53} &  \textbf{1.46} &  \textbf{1.45} &  \textbf{1.51} &  \textbf{1.41} &  \textbf{1.48} \\
Persistent MAE  &  2.16 &  2.24 &  2.13 &  1.67 &  1.87 &  2.12 &  1.76 &  1.68 &  1.77 &  2.33 &  2.32 &  2.54 &  2.05 \\
\\
Model 1 PCE $\tau = 0.25$   & \textbf{65.63} & \textbf{52.73} & \textbf{53.15} & \textbf{21.24} & \textbf{21.15} & \textbf{28.61} & \textbf{17.61} & \textbf{21.63} & \textbf{22.19} & \textbf{38.89} & \textbf{45.87} & \textbf{41.35} & \textbf{35.84} \\
Model 2 PCE $\tau = 0.25$   & 85.51 & 65.99 & 74.58 & 32.76 & 24.72 & 34.72 & 23.31 & 27.18 & 34.74 & 52.16 & 61.18 & 64.31 & 48.43 \\
Persistent PCE $\tau = 0.25$&115.36 & 79.25 & 96.12 & 44.29 & 29.41 & 41.84 & 28.99 & 33.73 & 47.39 & 65.24 & 74.31 & 87.26 & 61.93 \\
\\
Model 1 PCE $\tau = 0.5$    & 55.93 & 75.21 & 72.51 & 37.57 & 28.81 & 42.21 & 27.11 & 26.66 & 29.16 & 62.25 & 66.66 & 55.79 & 48.32 \\
Model 2 PCE $\tau = 0.5$    & \textbf{53.17}& \textbf{58.42}& \textbf{49.17}& \textbf{31.25}& \textbf{25.54}& \textbf{28.15}& \textbf{21.79} & \textbf{18.89} & \textbf{21.99} & \textbf{37.14} & \textbf{41.94} & \textbf{42.36} & \textbf{35.82} \\
Persistent PCE $\tau = 0.5$ & 58.78 & 91.99 & 95.85 & 44.89 & 32.16 & 56.25 & 33.43 & 34.43 & 37.34 & 87.35 & 91.37 & 69.22 & 61.09 \\
\\
Model 1 PCE $\tau = 0.75$   &28.36           &28.62           &41.13           &19.75           & 14.54         & 21.47 & 17.11 & 17.93 & 24.56 & 26.51 & 29.88 &23.18 & 24.42 \\
Model 2 PCE $\tau = 0.75$   & \textbf{27.44} & \textbf{24.54} & \textbf{41.91} & \textbf{19.23} &\textbf{12.73} & 21.66 &\textbf{16.74} & 18.11 &\textbf{21.71} & 27.78 &\textbf{28.36} & 25.426 &\textbf{23.80} \\
Persistent PCE $\tau = 0.75$&29.27           &32.69           &41.34           &21.26           & 16.35         &\textbf{21.29} & 17.27 &\textbf{17.76} & 28.41 &\textbf{25.22} & 31.41 & \textbf{21.75} & 25.34 \\
\\
    \multicolumn{14}{l}{\textbf{Gr\"unow}}  \\
Model 1 RMSE                &  1.72 &  1.74 &  \textbf{1.72} &  1.78 &  1.75 &  1.81 &  1.78 &  1.74 &  1.67 &  1.73 &  1.73 &  \textbf{1.62 }&  1.73 \\
Model 2 RMSE                &  \textbf{1.68}&  \textbf{1.71} &  1.73 & \textbf{ 1.76}&  \textbf{1.69}& \textbf{ 1.76}&  \textbf{1.74} &  \textbf{1.76 }&  \textbf{1.64} &  \textbf{1.73} & \textbf{1.71}&  \textbf{1.62}&  \textbf{1.71} \\
Persistent RMSE             &  2.46 &  2.45 &  2.29 &  2.12  &  2.14 &  2.11 &  2.18 &  2.18 &  2.11 &  2.72 &  2.41 &  2.72 &  2.32 \\
\\
Model 1 MAE                 &  1.51 &  1.56 &  \textbf{1.47}&  1.56 &  1.54 &  1.61 &  1.55 &  \textbf{1.52} &  1.49 &  1.54 &  1.56 &  1.45 &  1.53 \\
Model 2 MAE                 &  \textbf{1.46}&  \textbf{1.53}&  1.47 &  \textbf{1.56}&  \textbf{1.48}&  \textbf{1.56} & \textbf{1.51} &  1.53 &  \textbf{1.44} &  \textbf{1.54} &  \textbf{1.54}&  \textbf{1.44} &  \textbf{1.51} \\
Persistent MAE              &  2.34 &  2.32 &  2.16 &  2.15 &  1.94 &  1.89 &  1.95 &  2.16 &  2.11 &  2.65 &  2.33 &  2.66 &  2.22 \\
\\
Model 1 PCE $\tau = 0.25$   & \textbf{73.21} & \textbf{72.88} & \textbf{61.57} & \textbf{42.41} & \textbf{26.39} & \textbf{33.25} & \textbf{26.25} & \textbf{39.76} & \textbf{38.67} & \textbf{41.15} & \textbf{71.54} & \textbf{46.91} & \textbf{47.83} \\
Model 2 PCE $\tau = 0.25$   & 98.71 & 92.17 & 87.88 & 61.15 & 31.62 & 38.66 & 34.94 & 56.26 & 52.56 & 57.79 & 84.82 & 68.49 & 63.75 \\
Persistent PCE $\tau = 0.25$&124.18 &111.46 &114.18 & 77.71 & 36.85 & 44.16 & 43.63 & 72.76 & 66.46 & 74.44 & 99.19 & 91.19 & 79.68 \\
\\
Model 1 PCE $\tau = 0.5$    & 93.17 &111.29 & 85.57 & 55.87 & 31.18 & 47.26 & 31.24 & 54.41 & 54.41 & 87.15 & 83.49 & 66.86 & 66.83 \\
Model 2 PCE $\tau = 0.5$    & \textbf{83.39} & \textbf{82.42} & \textbf{63.16} & \textbf{41.87} & \textbf{25.71} & \textbf{32.71} & \textbf{24.51} & \textbf{41.41} & \textbf{42.15} & \textbf{57.21} & \textbf{58.33} & \textbf{55.74} & \textbf{50.72} \\
Persistent PCE $\tau = 0.5$ &112.75 &118.16 &118.17 & 71.86 & 36.66 & 61.81 & 37.96 & 68.41 & 66.76 &116.91 &118.65 & 77.97 & 83.84 \\
\\
Model 1 PCE $\tau = 0.75$   & 33.16 & 41.77 & 34.38 & 23.17 & 19.68 & 23.63 & 22.12 & 28.56 & 28.94 & 32.98 & 33.62 & 21.95 & 28.66 \\
Model 2 PCE $\tau = 0.75$   & 35.18 & 42.24 &\textbf{31.12} &\textbf{21.73} & 21.68 &\textbf{21.54} &\textbf{21.97} &\textbf{25.79} & 33.63 &\textbf{32.61} &\textbf{31.49} &\textbf{21.14} &\textbf{28.34} \\
Persistent PCE $\tau = 0.75$&\textbf{31.94} &\textbf{39.29} & 38.74 & 24.41 &\textbf{17.69} & 25.71 & 23.17 & 31.32 &\textbf{24.25} & 33.36 & 35.76 & 22.76 & 29.03 \\

\label{table:pred2}
\end{tabular} 
\end{center}
\end{tiny}
\end{sidewaystable}

\label{lastpage}

\end{document}